\documentclass[10pt,conference]{IEEEtran}

\usepackage[ruled,lined,linesnumbered,vlined]{algorithm2e}

\SetCommentSty{mycommfont}
\DontPrintSemicolon %

\SetKwProg{Fn}{Function}{:}{}

\usepackage{amsmath}
\usepackage{amsfonts}
\usepackage{amsthm}
\usepackage{balance}
\usepackage{enumitem}
\usepackage{graphicx}
\usepackage{listings}
\usepackage{multicol}
\usepackage{multirow}
\usepackage{subcaption}
\usepackage{url}
\usepackage{xspace}
\usepackage{xcolor}
\usepackage{tcolorbox}
\usepackage{colortbl}

\usepackage{tikz}
\usepackage{pgfplots}
\newcommand*\circled[1]{\tikz[baseline=(char.base)]{
    \node[shape=circle,draw,inner sep=0.5pt] (char) {\small#1};}}
\usetikzlibrary{shapes}
\usetikzlibrary{shapes.geometric}
\usetikzlibrary{arrows.meta, positioning}

\newcommand{\eg}{\mbox{\textit{e.g.}}\xspace}
\newcommand{\etal}{\mbox{\textit{et al.}}\xspace}

\newcommand{\ie}{\mbox{\textit{i.e.}}\xspace}

\newcommand{\wrt}{\mbox{w.r.t.}\xspace}

\definecolor{BlindColorTolOne}{HTML}{332288}
\definecolor{BlindColorTolTwo}{HTML}{117733} %
\definecolor{BlindColorTolThree}{HTML}{44AA99}
\definecolor{BlindColorTolFour}{HTML}{88CCEE}
\definecolor{BlindColorTolFive}{HTML}{DDCC77}
\definecolor{BlindColorTolSix}{HTML}{CC6677} %
\definecolor{BlindColorTolSeven}{HTML}{AA4499}
\definecolor{BlindColorTolEight}{HTML}{882255}

\definecolor{BlindColorWongOne}{HTML}{000000} %
\definecolor{BlindColorWongTwo}{HTML}{E69F00}
\definecolor{BlindColorWongThree}{HTML}{56B4E9}
\definecolor{BlindColorWongFour}{HTML}{009E73}
\definecolor{BlindColorWongFive}{HTML}{F0E442}
\definecolor{BlindColorWongSix}{HTML}{0072B2} %
\definecolor{BlindColorWongSeven}{HTML}{D55E00}
\definecolor{BlindColorWongEight}{HTML}{CC79A7}

\definecolor{mygreen}{HTML}{02818a}

\mathchardef\mhyphen="2D

\newcounter{FindingCounter}

\SetKwData{AlgTrue}{true}
\SetKwData{AlgFalse}{false}
\SetKw{AlgBreak}{break}
\SetKw{AlgContinue}{continue}

\newcounter{myUniqueIdCounter}

\makeatletter
\newcommand{\myGetOrAssignID}[1]{%
  \ifcsname myMap@#1\endcsname%
    \csname myMap@#1\endcsname%
  \else%
    \stepcounter{myUniqueIdCounter}%
    \expandafter\xdef\csname myMap@#1\endcsname{\themyUniqueIdCounter}%
    \themyUniqueIdCounter%
  \fi%
}
\makeatother

\newcommand{\anonymizedId}[1]{\ifx\useAnonymizedId\undefined%
  #1%
\else%
  \myGetOrAssignID{#1}%
\fi}

\SetCommentSty{mycommfont}

\newcommand*\filledcircled[1]{\tikz[baseline=(char.base)]{
    \node[shape=circle,fill=black,inner sep=1pt] (char) {\small\textcolor{white}{#1}};}}

\newcommand{\mytightparagraph}[1]{
  \vspace*{0.06cm}
  \noindent \textit{\textbf{#1}}\
}

\newcommand{\proj}{\textsc{Proj}\xspace}

\newcommand{\projCold}{\textsc{Proj-cold}\xspace}

\newcommand{\llm}{LLM\xspace}
\newcommand{\llms}{LLMs\xspace}

\newcommand{\sprs}{SPRs\xspace}
\newcommand{\agr}{AGR\xspace}
\newcommand{\agrs}{AGRs\xspace}
\newcommand{\creduce}{C-Reduce\xspace}
\newcommand{\perses}{Perses\xspace}
\newcommand{\vulcan}{Vulcan\xspace}

\newcommand{\lpr}{LPR\xspace}

\newcommand{\ddmin}{ddmin\xspace}

\newcommand{\languageC}{C\xspace}
\newcommand{\languageRust}{Rust\xspace}
\newcommand{\languageJS}{JavaScript\xspace}
\newcommand{\gcc}{GCC\xspace}
\newcommand{\llvm}{LLVM\xspace}
\newcommand{\clang}{Clang\xspace}

\newcommand{\rustc}{Rustc\xspace}

\newcommand{\javascriptcore}{JavaScriptCore\xspace}

\newcommand{\python}{Python\xspace}

\newcommand{\deepseekVfourflash}{DeepSeek-V4-Flash\xspace}
\newcommand{\mimoVTwoPointFive}{Mimo-V2.5\xspace}

\newcommand{\miniMaxMThree}{MiniMax-M3\xspace}

\newcommand{\miniSweAgent}{mini-SWE-agent\xspace}
\newcommand{\aider}{Aider\xspace}

\newcommand{\program}{\ensuremath{{P}}\xspace}
\newcommand{\programPrime}{\ensuremath{{P}^\prime}\xspace}
\newcommand{\programMin}{\ensuremath{{P}_{\textit{min}}}\xspace}
\newcommand{\programPrev}{\ensuremath{{P}_{\textit{prev}}}\xspace}
\newcommand{\programCand}{\ensuremath{{P}^\ast}\xspace}
\newcommand{\programPath}{\ensuremath{{P}_{\textit{path}}}\xspace}
\newcommand{\property}{\ensuremath{\psi}\xspace}

\newcommand{\transformation}{\ensuremath{\tau}\xspace}
\newcommand{\transformations}{\ensuremath{\mathcal{T}}\xspace}
\newcommand{\metadata}{\ensuremath{\mathcal{M}}\xspace}

\newcommand{\counter}{\ensuremath{cnt}\xspace}

\newcommand{\benchmarkC}{\texttt{Benchmark-C}\xspace}
\newcommand{\benchmarkRust}{\texttt{Benchmark-Rust}\xspace}
\newcommand{\benchmarkJavaScript}{\texttt{Benchmark-JS}\xspace}
\newcommand{\benchmarkCTrain}{\texttt{Benchmark-C-Train}\xspace}
\newcommand{\benchmarkCTest}{\texttt{Benchmark-C-Test}\xspace}

\newcommand{\maxTurnsFirstIteration}{60\xspace}
\newcommand{\maxTurnsSecondIteration}{50\xspace}
\newcommand{\maxTurnsThirdIteration}{40\xspace}
\newcommand{\maxTurnsTotal}{150\xspace}

\newcommand{\numBenchmarksAll}{90\xspace}
\newcommand{\numBenchmarksC}{60\xspace}
\newcommand{\numBenchmarksRust}{20\xspace}
\newcommand{\numBenchmarksJavaScript}{10\xspace}

\newcommand{\astraAverageSizeC}{82.2\xspace}
\newcommand{\astraAverageSizeRust}{117.7\xspace}
\newcommand{\astraAverageSizeJavaScript}{23.3\xspace}

\newcommand{\astraImprovementOverCreduceC}{39.0\%\xspace}

\newcommand{\astraImprovementOverVulcanRust}{36.0\%\xspace}
\newcommand{\astraImprovementOverVulcanJavaScript}{38.9\%\xspace}

\newcommand{\astraImprovementOverBestBaselineC}{\astraImprovementOverCreduceC}
\newcommand{\astraImprovementOverBestBaselineRust}{\astraImprovementOverVulcanRust}
\newcommand{\astraImprovementOverBestBaselineJavaScript}{\astraImprovementOverVulcanJavaScript}

\newcommand{\astraImprovementOverVulcanTime}{77.2\%\xspace}
\newcommand{\astraImprovementOverLPRTime}{42.6\%\xspace}
\newcommand{\astraImprovementOverCreduceTime}{57.7\%\xspace}

\newcommand{\astraVsAstraColdLLMTurns}{33\%\xspace}
\newcommand{\astraVsAstraColdCost}{26\%\xspace}

\usepackage{cleveref} %

\Crefname{algocf}{Algorithm}{Algorithms}
\crefname{algocf}{Algorithm}{Algorithms}

\Crefname{algorithm}{Algorithm}{Algorithms}
\crefname{algorithm}{Algorithm}{Algorithms}

\crefname{appendix}{Appendix}{Appendices}
\Crefname{appendix}{Appendix}{Appendices}

\Crefname{figure}{Figure}{Figures}
\crefname{figure}{Figure}{Figures}

\crefname{listing}{Listing}{Listings}
\Crefname{listing}{Listing}{Listings}

\Crefname{table}{Table}{Tables}
\crefname{table}{Table}{Tables}

\crefname{thm}{Theorem}{Theorems}
\Crefname{thm}{Theorem}{Theorems}

\crefname{equation}{Equation}{Equations}
\Crefname{equation}{Equation}{Equations}

\crefformat{chapter}{\S~#2#1#3}
\crefmultiformat{chapter}{\S\S~#2#1#3}{ and~#2#1#3}{, #2#1#3}{, and~#2#1#3}

\crefformat{section}{\S~#2#1#3}
\crefmultiformat{section}{\S\S~#2#1#3}{ and~#2#1#3}{, #2#1#3}{, and~#2#1#3}

\usepackage{booktabs}

\begin{document}

\title{
	Semantic-aware and Self-improving Program Reduction via Agentic Large Language Models
}

\author{%
\parbox{\textwidth}{%
\parbox[t]{0.32\textwidth}{\centering
{\sublargesize Xintong Zhou}\\
{\normalsize \textit{University of Waterloo}\\
Waterloo, Canada\\
x27zhou@uwaterloo.ca}
}%
\hfill
\parbox[t]{0.32\textwidth}{\centering
{\sublargesize Hongxu Xu}\\
{\normalsize \textit{University of Waterloo}\\
Waterloo, Canada\\
hongxu.xu@uwaterloo.ca}
}%
\hfill
\parbox[t]{0.32\textwidth}{\centering
{\sublargesize Chunhao Liao}\\
{\normalsize \textit{University of Waterloo}\\
Waterloo, Canada\\
chunhao.liao@uwaterloo.ca}
}%
\par\vspace{1.5ex}%
\parbox[t]{0.32\textwidth}{\centering
{\sublargesize Puzhuo Liu}\\
{\normalsize \textit{Tsinghua University}\\
Beijing, China\\
liupz@mail.tsinghua.edu.cn}
}%
\hfill
\parbox[t]{0.32\textwidth}{\centering
{\sublargesize Yongqiang Tian}\\
{\normalsize \textit{Monash University}\\
Melbourne, Australia\\
yongqiang.tian@monash.edu}
}%
\hfill
\parbox[t]{0.32\textwidth}{\centering
{\sublargesize Chengnian Sun}\\
{\normalsize \textit{University of Waterloo}\\
Waterloo, Canada\\
cnsun@uwaterloo.ca}
}%
}%
}

\maketitle

\begin{abstract}
	Reducing bug-triggering programs to their
minimal essential form is a fundamental task
in debugging language processors such as compilers
and interpreters. Existing reduction techniques
are limited by their reliance on predefined,
syntax-driven transformations that lack semantic
understanding of the target program, and by
their inability to learn from past
reduction experiences.

We present a new approach that recasts program
reduction as an autonomous reasoning task
powered by agentic Large Language Models (\llms).
Instead of applying fixed transformation rules,
our method enables an \llm to analyze program
semantics, formulate reduction hypotheses, and
iteratively refine its approach based on execution
outcomes. Successful reduction experiences are
further distilled into reusable strategies,
allowing the system to continuously improve
over time.

We realize this approach in \proj, a framework
built around two collaborative components:
a \emph{reducer agent} that performs
semantic-aware, case-specific program reduction,
and a \emph{reflector agent} that extracts and
accumulates transferable reduction knowledge.
Extensive experiments on \numBenchmarksAll benchmarks
spanning three programming languages show that
\proj consistently produces smaller reduced programs
than all existing state-of-the-art reducers
while maintaining high efficiency.

\end{abstract}

\section{Introduction}
\label{sec:introduction}
Program reduction is an essential and widely applied
technique for debugging language processors,
such as compilers, interpreters, and debuggers~\cite{emi,
csmith,live-code-mutation,yarpgen,lidbury2015many}.
Given a program \program that satisfies a property \property
(\eg, \program triggers a bug in a language processor)
along with a property checker that verifies whether
a program satisfies \property,
program reduction aims to produce a minimized program
\programMin that still satisfies \property.
By removing bug-irrelevant fragments from \program,
program reduction helps developers
isolate the underlying defect, thereby streamlining
the debugging process.
The significance of program reduction is evidenced
by the fact that many language processor communities
mandate the bug-triggering program to be minimized before
submitting a bug report~\cite{gccbugreport,
llvmbugreport,jerrybugreport,cpythonreport}.

To address this critical problem,
many reduction techniques have been proposed
over the past decades~\cite{creduce,perses,vulcan,
xu2025boosting,trec,lpr}.
Existing techniques can be broadly divided
into two categories:
language-agnostic program reducers (\agrs)
and language-specific program reducers (\sprs).
\agrs, such as \perses~\cite{perses} and \vulcan~\cite{vulcan},
are designed to handle any programming language
with language-agnostic reduction strategies,
while \sprs, such as \creduce~\cite{creduce},
are tailored to particular programming languages
and leverage the language-specific features and
semantics to facilitate the reduction process.
Despite these efforts,
the effectiveness of existing reducers in practice
often falls short of expectations.
A telling piece of evidence is that
compiler developers frequently need to
manually ``further reduce'' the program
submitted in bug reports to assist the debugging
process~\cite{llvm-198755,gcc-125538}.
We attribute this gap to two fundamental limitations
of existing reducers.

\noindent
\textbf{Limitation 1: Lack of Semantic Awareness.}
Existing reducers suffer from a fundamental lack of
semantic awareness of the specific program and the
property it must preserve.
The inability of \agrs to exploit language-specific
semantics is self-evident.
However, even \sprs like \creduce are essentially static,
human-designed heuristic systems:
they apply a fixed set of reduction strategies
(\eg, removing unused functions, inlining macros)
in a predefined order, without any understanding of the
particular program being reduced or what its property entails.
As a result, many reduction opportunities that require
case-specific semantic reasoning are missed.
For example, recognizing that a particular computation is
redundant given the property, or that two code fragments can be
unified, demands an understanding of the program's semantics
in the context of the specific case, which a fixed,
program-agnostic rule set cannot provide.
Recent work begins to bring \llms to bear,
but only partially:
\lpr~\cite{lpr} invokes an \llm to apply a few fixed,
human-designed transformations (\eg, function inlining),
injecting some semantic knowledge yet remaining bound to a
predetermined repertoire rather than reasoning about the
specific program and property at hand.

\noindent
\textbf{Limitation 2: Lack of Self-Evolution.}
Existing reducers cannot evolve autonomously
over time: their repertoire of reduction rules is fixed at
design time and does not improve with use.
Augmenting a reducer with a new reduction strategy requires
domain experts to manually design, implement, and
validate it, while the reducer itself learns nothing
from the many reductions it performs.
Consequently, these tools cannot keep pace with the diversity
of reduction patterns encountered in practice:
every newly observed reduction pattern must be anticipated
and manually encoded by an expert, rather than being acquired
automatically from accumulated experiences.

\mytightparagraph{Agentic Program Reduction.}
In this paper, we propose \emph{agentic program reduction},
a new paradigm that reimagines the program reduction process
by leveraging the power of agentic \llms beyond
the passive usage of processing predefined transformations.
Our key insight is to treat program reduction as
an \emph{exploratory reasoning} process.
Indeed, a skilled developer does not follow a fixed checklist
of transformations, but instead analyzes the program,
forms a hypothesis about what can be removed,
attempts the edit, observes the outcome,
and adjusts the strategy accordingly.
\llms, as agentic reasoners, can perform exactly
this kind of iterative, feedback-driven problem solving.
By harnessing the \llm within a structured reduction
pipeline while granting it full autonomy over the process,
we unlock the semantic-aware reduction capabilities that
no fixed set of transformations can provide.

On the other hand, treating every reduction as a fresh
reasoning problem is itself imperfect: invoking the
\llm at each step is costly and non-deterministic,
even when the required edit is entirely routine. Many
reduction operations~\cite{latra,lpr},
in fact, follow \emph{general, recurring}
patterns,
such as inlining a forwarding wrapper,
that once discovered, could be reapplied mechanically
with no further reasoning, yet a purely agentic reducer
re-derives them from scratch on every program. We
therefore complement the reducer with a reflective
step: by allowing the \llm to \emph{reflect} on its
past reduction experiences and distill them into
reusable strategies, recurring patterns are captured
once and reapplied mechanically thereafter, with no
further reasoning. This is a new
paradigm in its own right: reduction knowledge that is
no longer hand-crafted by experts but
\emph{automatically accumulated from experience}, letting
the reducer grow as it is used.

\mytightparagraph{\proj.}
Based on these insights, we design and develop \proj,
an LLM-based agentic framework for program reduction.
\proj harnesses \llms' capabilities for program
reduction through two key components.
First, a \emph{reducer agent} performs semantic-aware
reduction in a ReAct loop~\cite{yao2022react}:
it analyzes the program and the property,
proposes a candidate reduction, checks whether
the candidate preserves the property, and iterates.
After the reduction process converges,
an offline \emph{reflector agent} automatically
codifies successful reduction experiences into generalized,
reusable reduction strategies by maintaining a
\emph{learned reducer}, which is
a deterministic program that encapsulates the learned
knowledge and can be directly invoked
in future reduction sessions without any \llm intervention.

We extensively evaluate \proj on \numBenchmarksAll benchmarks
across three programming languages,
\ie, \languageC, \languageRust, and \languageJS.
The results demonstrate the superior capability of \proj
in program reduction compared with existing reducers.
Specifically, \proj produces
\astraImprovementOverBestBaselineC,
\astraImprovementOverBestBaselineRust, and
\astraImprovementOverBestBaselineJavaScript smaller
results than the best baseline on \languageC,
\languageRust, and \languageJS benchmarks, respectively,
while keeping significantly higher efficiency.
Notably,
to the best of our knowledge,
on \languageC programs,
\proj is the first
reducer that outperforms \creduce,
the long-standing de facto standard
for \languageC program reduction.
More importantly, \proj continuously accumulates
reusable reduction strategies in its learned reducer,
improving its scalability while saving costs.

\mytightparagraph{Contributions.}
This paper makes the following contributions:
\begin{itemize}[leftmargin=*, topsep=0pt, itemsep=0pt, partopsep=0pt]
    \item We propose \emph{agentic program reduction},
    a new paradigm that reimagines
    program reduction as an exploratory reasoning process,
    and harnesses the power of agentic \llms to
    assist this process.

    \item We design and develop \proj, an \llm-based
    agentic framework to perform language-agnostic and
    semantic-aware program reduction.

    \item We comprehensively evaluate \proj on \numBenchmarksAll
    benchmarks across three programming languages,
    \ie, \languageC, \languageRust, and \languageJS.
    The results show the superior performance of \proj
    over existing approaches.

\end{itemize}

\section{Motivating Example}
\label{sec:motivating-example}
\begin{figure*}[!h]
  \newlength{\examplesubfigheight}
  \setlength{\examplesubfigheight}{8cm}
  \hspace{1.4em}%
  \begin{subfigure}[b]{0.3\textwidth}
      \begin{minipage}[t][\examplesubfigheight][t]{\linewidth}
\begin{lstlisting}[
  firstnumber=1,
  numbers=left,
  numbersep=6pt,
]
struct a {
  signed b : 18;
  unsigned c;
  unsigned d;
  signed e;
  signed f;
};
struct a g;
struct a h;
int i() {
struct a j =
  {24, 6738, 165};
  int k = h.c;
  j.b = k;
  int l = h.f && j.b;
  int m = j.b;
  signed n =
    m | -(j.d && j.b) - l;
  g.e = n;
}
int main() {}
\end{lstlisting}
          \vfill
          \captionsetup{singlelinecheck=false, margin={-1.5em, 0.1\linewidth}}
          \vspace{1.2em}
          \caption{
              The result of \perses (\textbf{109 tokens}).
          }
          \label{subfig:example:perses-result}
      \end{minipage}
  \end{subfigure}
  \hspace{1.5em}%
  \begin{subfigure}[b]{0.29\textwidth}
      \begin{minipage}[t][\examplesubfigheight][t]{\linewidth}
\begin{lstlisting}[firstnumber=1, numbers=left, numbersep=6pt, escapechar=@]
struct a {
  signed b : 18;
  @\colorbox{blue!15}{unsigned c;}@
  @\colorbox{green!15}{signed e;}@
  @\colorbox{red!15}{signed f;}@
} g, h;
int i() {
  struct a j;
  j.b = @\colorbox{blue!15}{h.c}@;
  int l = @\colorbox{red!15}{h.f}@ && j.b;
  @\colorbox{green!15}{g.e}@ = j.b | -!!j.b - l;
}
int main() {}
\end{lstlisting}
          \vfill
          \captionsetup{singlelinecheck=false, margin={-1.5em, 0.1\linewidth}}
          \vspace{-0.2em}
          \caption{
              An intermediate result progressed by
              \proj. In detail,
              \proj inlines single use variable
              \texttt{k}, \texttt{n} and \texttt{m}
              (\eg, \texttt{k=h.c;j.b=k;} $\rightarrow$ \texttt{j.b=h.c;}),
              and simplifies \texttt{-(j.d\&\&j.b)} to \texttt{-!!j.b}
              since \texttt{j.d} is always non-zero (\texttt{165}),
              and subsequently remove the initialization of \texttt{j}
              and the field \texttt{d}.
          }
          \label{subfig:example:intermediate-result-1}
      \end{minipage}
  \end{subfigure}
  \hspace{2.4em}%
  \begin{minipage}[t][\examplesubfigheight][t]{0.32\textwidth}
    \begin{subfigure}[t]{\linewidth}
\begin{lstlisting}[firstnumber=1, numbers=left, numbersep=6pt, escapechar=@]
unsigned @\colorbox{blue!15}{h\_c}@,@\colorbox{green!15}{g\_e}@,@\colorbox{red!15}{h\_f}@;
struct a { signed b : 18; };
int i() {
  struct a j;
  j.b = @\colorbox{blue!15}{h\_c}@;
  int l = @\colorbox{red!15}{h\_f}@ && j.b;
  @\colorbox{green!15}{g\_e}@ = j.b | -!!j.b - l;
}
int main() {}
\end{lstlisting}
      \captionsetup{singlelinecheck=false, margin={-1.5em, 0.1\linewidth}}
      \vspace{-0.5em}
      \caption{
          The intermediate result after
          hoisting struct fields and
          inlining single-use variables.
      }
      \label{subfig:example:intermediate-result-2}
    \end{subfigure}

    \vspace{0.2em}
    \begin{subfigure}[t]{\linewidth}
\begin{lstlisting}[firstnumber=1, numbers=left, numbersep=6pt, escapechar=@]
int t;
int main() {
  struct {
    signed b : 18;
  } j;
  j.b = t;
  int l = t && j.b;
  t = j.b | -!!j.b - l;
}
\end{lstlisting}
      \captionsetup{singlelinecheck=false, margin={-0.8em, 0.1\linewidth}}
      \vspace{-0.7em}
      \caption{
          Final result of \proj (\textbf{49 tokens}).
      }
      \label{subfig:example:astra-final-result}
    \end{subfigure}
  \end{minipage}
  \vspace{0.3em}
  \caption{
      A motivating example of \proj's reduction process. \perses+\proj produces 49 tokens,
      which is much smaller than 68 tokens produced by \creduce
      and 82 tokens produced by \lpr.
  }
  \label{fig:motivating_example}

  \vspace{-1em}
\end{figure*}

We illustrate \proj's key ideas on a real-world bug in the
\clang compiler~\cite{llvm27747}.
\cref{subfig:example:perses-result} shows the program after
\perses reduces it to a fixpoint of 109 tokens.
As a purely
syntactic reducer, \perses leaves behind reductions that
require reasoning about the program's semantics.
\proj then continues to perform semantic-aware reductions
with its reducer agent (\cref{subsec:reducer-agent}).
\cref{subfig:example:intermediate-result-1} shows an intermediate
result.
\proj
inlines the single-use temporaries \texttt{k},
\texttt{m}, and \texttt{n}
(\eg, replacing \texttt{k=h.c;j.b=k;} with \texttt{j.b=h.c;}),
and, more subtly, recognizes that \texttt{j.d} always
holds the non-zero constant \texttt{165}, so the conjunction
\texttt{j.d\,\&\&\,j.b} reduces to a Boolean test of
\texttt{j.b}.
This simplifies \texttt{-(j.d\,\&\&\,j.b)} to \texttt{-!!j.b}
and renders the field \texttt{d} and the initializer of
\texttt{j} dead, both are subsequently removed.
Each such edit requires an understanding of the program's
semantics in the context of the specific case.

Reasoning at a larger scope, the agent then restructures the
program's struct types (\cref{subfig:example:intermediate-result-2}).
It observes that the structs such as \texttt{g} and
\texttt{h}, instances of the struct type \texttt{a},
are referenced \emph{only} through individual
fields (\eg, \texttt{g.e} and \texttt{h.c},
highlighted in \cref{subfig:example:intermediate-result-1}).
It therefore attempts to dissolve the structs, hoisting each
exclusively-accessed field into a scalar global
(\eg, \texttt{g.e}$\rightarrow$\texttt{g\_e})
and stripping these fields from \texttt{a}.
Consequently, fields \texttt{c}, \texttt{e}, and \texttt{f}
are removed from \texttt{a}, leaving only the bit-field
\texttt{b} in the struct,
which is essential to trigger the bug.
Furthermore, \proj observes that
the three scalar globals carry no bug-relevant identity and
coalesces them into a single global \texttt{t}. It also
inlines the helper function \texttt{i} into \texttt{main}
and drops the redundant struct tag \texttt{a}.
Finally, what remains is the core of the bug
shown in \cref{subfig:example:astra-final-result}:
a single 18-bit bit-field and the expression
\texttt{j.b\,|\,-!!j.b\,-\,l} that triggers it.
At 49 tokens,
the final result is less than half the size of
\perses, and much smaller than
\creduce (68) and \lpr (82).

Moreover, the value of these reductions extends beyond this
single case. After the session completes,
\proj's reflector agent (\cref{subsec:reflector-agent})
reviews the accepted reductions
and distills them into generalized,
reusable strategies in the learned reducer.
For example, from this case,
the reflector extracts a strategy that hoists
the exclusively-accessed field of a global struct
into a scalar variable.
Thereafter the learned reducer applies these strategies
automatically, with no \llm intervention,
to any future program that exhibits the same pattern.

\section{Approach}
\label{sec:approach}
This section presents the workflow of \proj
with a detailed explanation of each component.

\subsection{Workflow}
\label{subsec:workflow}

\begin{figure*}[!htbp]
    \centering
    \includegraphics[width=0.98\textwidth]{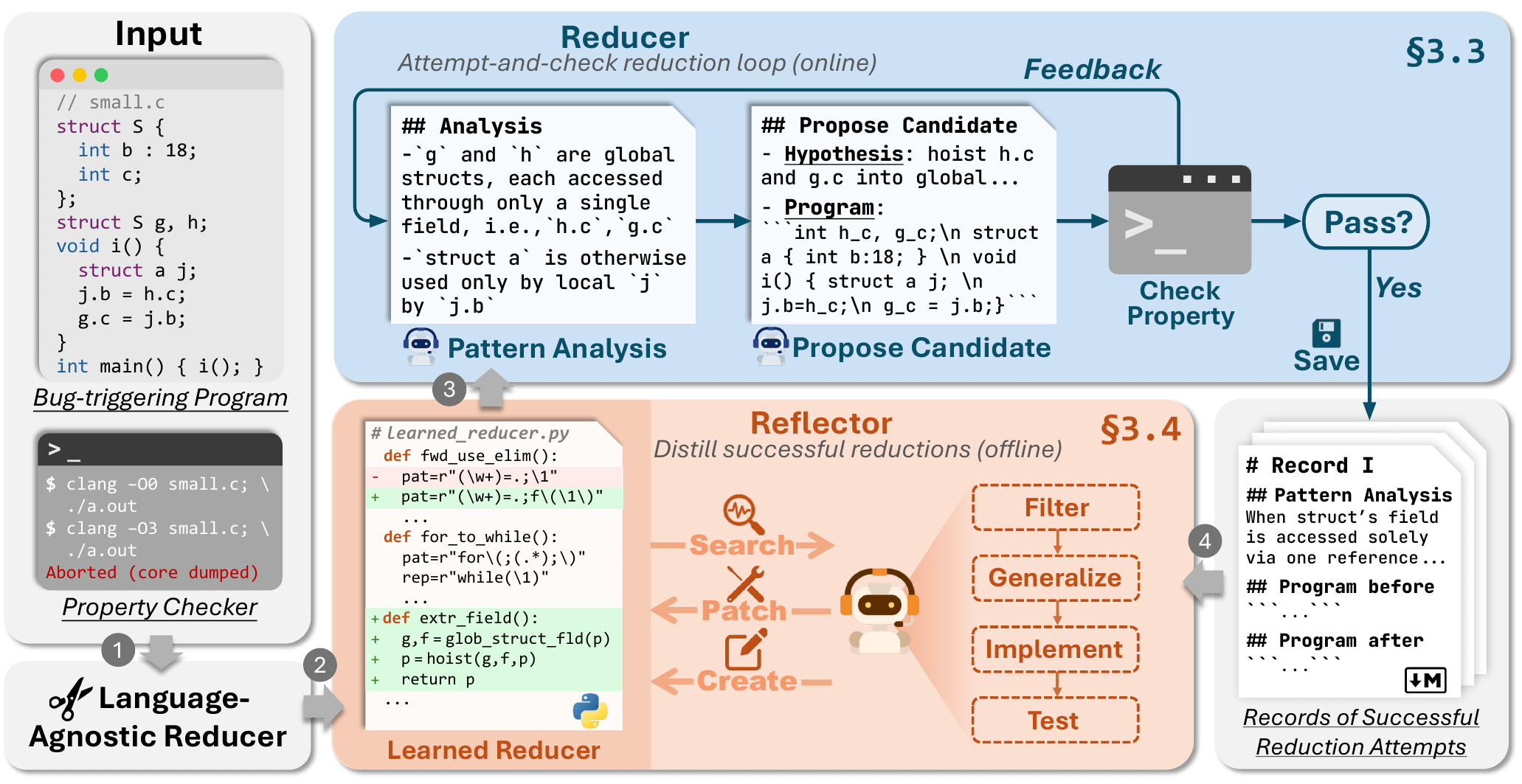}
    \caption{The workflow of \proj.
    }
    \label{fig:workflow}
\end{figure*}

The workflow of \proj is shown in \cref{fig:workflow}.
Given a bug-triggering program and a property checker that verifies
whether a program satisfies the property (\ie, triggers the bug),
\proj first invokes a language-agnostic reducer (\agr)
to reduce the program (\filledcircled{1}).
This design choice aligns with prior work,
such as \vulcan~\cite{vulcan} and \lpr~\cite{lpr},
where an \agr is capable of quickly reducing a large input into
manageable scale, enabling the subsequent fine-grained reduction.
Following prior work~\cite{vulcan,lpr},
\proj uses \perses~\cite{perses} as the default \agr,
though its design is not tied to any specific \agr and can
readily accommodate alternatives.
After the initial reduction, the program is passed to
the \emph{learned reducer} for further reduction (\filledcircled{2}).
The learned reducer is a deterministic reduction program
created and evolved by the \emph{reflector agent},
containing an increasing set of reduction strategies
distilled from historical reduction experiences.
After passing through the learned reducer,
the program is handed to the \emph{reducer agent} (\filledcircled{3}),
which performs semantic-aware reduction in a ReAct loop~\cite{yao2022react}:
it analyzes the program and the property,
hypothesizes the next reduction opportunity,
then proposes a candidate,
and checks whether the candidate satisfies the property.
After the online reduction process is completed,
the result is returned to the user,
and the reflector agent is invoked to perform offline distillation (\filledcircled{4}).
It examines the successful reduction attempts and distills them into
generalizable reduction strategies in the
learned reducer, which can be directly
invoked in future reduction
sessions without any \llm intervention.

\subsection{Execution Harness and Tool Interface}
\label{subsec:harness}

Each of \proj's two agents pairs the \llm
with a runtime harness that
turns its reasoning into a disciplined reduction process.
The harness owns all durable state, \eg, the current
working program, the best property-preserving program found
so far, and the learned reducer's strategy pool,
and drives every property check.
The \llm contributes reasoning; the harness contributes
memory, execution, and ground truth.

The harness exposes this state through a small, bounded
\emph{tool interface}, summarized in \cref{tab:tools}.
The \llm observes state and effects change \emph{solely}
by emitting tool calls, which the harness interprets,
executes, and answers with structured results that form the
agent's next observation. The tools thus act as a strict
control boundary: the model proposes through them, and the
harness disposes. No proposal is taken on trust: a
reduction candidate is adopted only after the harness runs
the property checker against it, and a distilled strategy
enters the learned reducer only after it clears a
verification pipeline. Every \llm decision is, in this way,
ratified by an executable check before it can alter the
system. This cleanly separates \emph{what} to reduce or distill,
left to the models' reasoning, from \emph{whether} a change
is sound, decided by deterministic machinery.
We describe each agent in detail in the following sections.

\begin{table}[t]
  \centering
  \caption{
    Tools exposed to \proj's two LLM agents.
  }
  \label{tab:tools}
  \scriptsize
  \renewcommand{\arraystretch}{0.95}
  \begin{tabular}{c@{\hspace{4pt}}|@{\hspace{3pt}}l@{\hspace{7pt}}p{5.3cm}@{}}
    \toprule
    & \textbf{Tool} & \textbf{Description} \\
    \midrule
    \multirow{6}{*}{\rotatebox[origin=c]{90}{\scriptsize\textbf{Reducer}}}
      & \texttt{read\_program}      & Return the current program and its token count. \\
      & \texttt{read\_best}         & Return the smallest oracle-passing program so far. \\
      & \texttt{read\_strategies}   & List the learned reducer's strategies. \\
      & \texttt{propose\_candidate} & Submit \& validate a candidate; update best on accept.\, \\
      & \texttt{revert\_to\_best}   & Reset the working program to the best-so-far. \\
      & \texttt{declare\_done}      & Terminate the reduction loop. \\
    \midrule
    \multirow{8}{*}{\rotatebox[origin=c]{90}{\scriptsize\textbf{Reflector}}}
      & \texttt{read\_records}        & Return reduction records with descriptions. \\
      & \texttt{read\_before}       & Return the program before reduction. \\
      & \texttt{read\_after}          & Return the program after reduction. \\
      & \texttt{read\_diff}           & Return a unified diff of original vs.\ final. \\
      & \texttt{list\_strategies}     & List descriptions of all learned strategies. \\
      & \texttt{read\_strategy}       & Return the source and fixtures of a strategy. \\
      & \texttt{propose\_action}      & Submit a strategy; stage and run verification. \\
      & \texttt{commit}               & Land a staged strategy into the learned reducer. \\
      & \texttt{declare\_done}        & Terminate the reflection loop. \\
    \bottomrule
  \end{tabular}
\end{table}

\subsection{Reducer Agent}
\label{subsec:reducer-agent}

The reducer agent is the core component of \proj that
performs the reduction process.
It receives a program pruned by the \agr and the learned reducer,
along with the property checker \property,
and further reduces the program with respect to the property.
Rather than following a fixed repertoire of transformations,
the reducer agent approaches reduction as an exploratory
reasoning process.
\cref{alg:reducer-agent} shows the workflow of the reducer agent.

\mytightparagraph{Two-Level Loop Structure.}
The reducer agent operates in two nested loops.
The outer loop (\cref{line:reducer:outer-loop})
runs up to $I_{\max}$ \emph{iterations},
where each iteration corresponds to a fresh \llm conversation
session with a clean context window
(\cref{line:reducer:new-session}).
Within each iteration, the inner loop
(\cref{line:reducer:inner-loop}-\ref{line:reducer:inner-loop-end})
executes up to $A_{\max}$ \emph{attempts} of the
propose-check-iterate cycle described below.
This two-level structure separates local planning
from global exploration.
Within an iteration, a continuous context lets the agent use recent
accepted and rejected attempts to plan coherent follow-up reductions.
Across iterations, however,
accumulated context may bias the agent
toward outdated hypotheses or
previously unsuccessful edit paths.
Starting a fresh session from the current best program
periodically refreshes the agent's perspective,
encouraging alternative reduction
opportunities while preserving all validated progress.
The agent may re-examine and analyze the program and the
property at the beginning of each session.
This preliminary analysis then equips the agent with
the context to make informed reduction decisions,
rather than resorting to blind trial and error.
We verify this two-level loop design in \cref{subsec:ablation-study}.

\begin{algorithm}[h!]
    \footnotesize
    \DontPrintSemicolon
    \SetKwInput{KwData}{Input}
    \SetKwInput{KwResult}{Output}
    \SetKwFunction{ProposeCandidate}{propose\_candidate}
    \SetKwFunction{NewSession}{new\_session}
    \SetKwFunction{Analyze}{Analyze}
    \SetKwFunction{DeclareDone}{declare\_done}
    \SetKwFunction{Revert}{revert\_to\_best}
    \SetKwFunction{LLM}{call\_llm}
    \SetKwFunction{NextToolCall}{next\_tool\_call}
    \SetKwFunction{Check}{Check}

    \caption{
        The workflow of the reducer agent.
    }
    \label{alg:reducer-agent}

    \KwData{$\program$: the input program to be reduced.}
    \KwData{$\property$: the property to be preserved.}
    \KwResult{$\programMin$: the minimized program that satisfies $\property$.}
    \KwResult{$\mathcal{R}$: the records of successful reductions.}

    \BlankLine
    $\programMin \leftarrow \program$,\
    $\mathcal{R} \leftarrow \emptyset$

    \For{$\textit{iter} \leftarrow 1$ \KwTo $I_{\max}$}{ \label{line:reducer:outer-loop}

        $\programPrime \leftarrow \programMin$\;

        $P_{\textit{path}} \leftarrow \programMin$,\
        $\counter \leftarrow 0$\; \label{line:reducer:init-counter}

        \NewSession{$\programMin, \property$} \tcc*[f]{\footnotesize start a fresh \llm session}\label{line:reducer:new-session}

        \For(\tcc*[f]{\footnotesize propose-check-iterate}){$j \leftarrow 1$ \KwTo $A_{\max}$}{ \label{line:reducer:inner-loop}

            $t \leftarrow$ \NextToolCall{}\label{line:reducer:next-tool-call}

            \If{$t = {}$\ProposeCandidate{$\programCand, \metadata$}}{ \label{line:reducer:propose-candidate}
                \If{$\property(\programCand)$}{ \label{line:reducer:check}
                    \uIf{$|\programCand| < |\programMin|$}{
                        $\programMin \leftarrow \programCand$,\
                        $\counter \leftarrow 0$\; \label{line:reducer:update-program-min}
                        $\mathcal{R} \leftarrow \mathcal{R} \cup
                        \{(P_{\textit{path}}, \programCand, \metadata)\}$\tcc*[f]{\footnotesize record} \label{line:reducer:record}
                    }
                    \Else{
                        $\counter \leftarrow \counter + 1$\; \label{line:reducer:increment-counter}
                    }
                    $P_{\textit{path}} \leftarrow \programCand$\;
                }
            }
            \If{$t = {}$\Revert{} $\mid\mid$ $\counter = N$}{
                $P_{\textit{path}} \leftarrow \programMin$,\
                $\counter \leftarrow 0$\tcc*[f]{\footnotesize revert to \programMin}\label{line:reducer:revert-action}
            }

            \lIf{$t = {}$\DeclareDone{}}{\AlgBreak} \label{line:reducer:declare-done}

        }
        \lIf{$\programMin = \programPrime$}{\AlgBreak\tcc*[f]{\footnotesize stop if no progress}}\label{line:reducer:converge} \label{line:reducer:inner-loop-end}
    }
    \Return $(\programMin, \mathcal{R})$\;
\end{algorithm}

\mytightparagraph{Propose-Check-Iterate.}
The reducer agent follows a structured propose-check-iterate loop
(\cref{line:reducer:inner-loop}-\ref{line:reducer:inner-loop-end}).
At each step,
the agent emits a tool call given its current
session context (\cref{line:reducer:next-tool-call}),
and the harness carries it out:
it may submit a candidate for checking
(\cref{line:reducer:propose-candidate}),
reverts to the best program (\cref{line:reducer:revert-action}),
or declares the reduction complete (\cref{line:reducer:declare-done}).
For candidate proposal,
the agent is required to first articulate
the \emph{rationale} for its next action in natural language,
for example,
``simplify the nested conditional into a single branch''
or ``inline the call to \texttt{foo()} since it has
a single call site''.
Only after stating this rationale does the agent
produce the concrete candidate program.
The \texttt{propose\_candidate} tool then takes
two arguments:
the candidate program \programCand and the
natural-language description \metadata
(\cref{line:reducer:propose-candidate}).
The harness receives the tool call
and then executes the property checker (\cref{line:reducer:check})
on the candidate and returns the result
(\ie, pass or fail) to the agent as feedback.
When a candidate passes the property check,
the reduction is accepted and the current program is updated
(\cref{line:reducer:update-program-min}).
The harness simultaneously records a triple
$(\programPrev, \programCand, \metadata)$
(\cref{line:reducer:record}),
capturing three key pieces of information:
the program before and after the reduction
($\programPrev$ and $\programCand$),
and the rationale for the reduction in natural language
($\metadata$).
These records are not consumed by the reducer agent itself;
instead, they are forwarded to the reflector agent
after the reduction session concludes,
serving as the raw material for strategy distillation
(\cref{subsec:reflector-agent}).
Requiring the agent to externalize its reasoning before
each action serves a dual purpose:
it discourages impulsive, low-quality attempts
during reduction, and it produces structured annotations
that make downstream distillation more effective.

\mytightparagraph{Two Reduction Modes.}
The reducer agent is instructed to
perform reduction in two complementary modes:

\noindent
\circled{1}~\emph{Default mode} (primary).
The agent directly reduces the program size by removing
or simplifying program fragments.
This mode aligns with the primary objective of
program reduction and accounts for the majority
of the agent's actions.
To prevent the agent from making erratic changes,
we instruct it to produce structured, logically motivated edits
targeting a specific, well-identified reduction opportunity.

\noindent
\circled{2}~\emph{Exploration mode} (tactical).
In certain situations, a transformation that
\emph{temporarily increases} program size can unlock
subsequent reduction opportunities that ultimately
yield a smaller program~\cite{lpr}.
For example, unrolling a loop temporarily
increases the token count
but the resulting straight-line code often exposes
redundant statements that can then be removed.
The agent is permitted to make such
expansionary moves when it can justify the potential payoff.
To prevent the agent from diverging on an unproductive path,
we impose an \emph{exploration budget}:
the harness maintains the current best
(\ie, smallest) program $\programMin$ alongside
the program on the current exploration path $\programPath$,
as well as a counter
$\counter$ of consecutive rounds without improvement
(\cref{line:reducer:increment-counter}).
If the exploration path fails to improve upon $\programMin$
within $N$ consecutive rounds (\eg, $N=5$ in our settings),
the harness reverts to $\programMin$
and instructs the agent to abandon the current direction
(\cref{line:reducer:revert-action}).

\mytightparagraph{Termination.}
The inner loop of a single iteration ends when
the agent declares that no further reduction
opportunity remains (\cref{line:reducer:declare-done}),
or the attempt count reaches $A_{\max}$.
The outer loop terminates when the iteration count
reaches $I_{\max}$, or when an iteration produces
no progress over the previous one
(\cref{line:reducer:converge}),
indicating that a fresh context no longer
helps and the reduction has converged.

\subsection{Reflector Agent}
\label{subsec:reflector-agent}

After the reduction process is completed, the reflector agent
performs an \emph{offline distillation} over the completed case.
It examines the successful reduction operations accepted by
the property checker, identifies recurring semantic patterns
behind them, and distills these patterns into generalized,
reusable reduction strategies.
Each strategy is then materialized as an \emph{executable pass}
in a standalone \python program called the \emph{learned reducer},
which can be directly invoked in future reduction sessions
without any \llm intervention.
The reflector agent is the key mechanism through which \proj
becomes self-improving:
as \proj processes more cases, the reflector agent continuously
accumulates reduction strategies into the learned reducer,
enabling \proj to handle future cases more stably with
less \llm intervention.

\mytightparagraph{Learned Reducer.}
Recent agent frameworks often improve by accumulating experience
as natural-language reflections~\cite{shinn2023reflexion}
or reusable executable skills~\cite{wang2023voyager}.
\proj follows the same principle but specializes it for program
reduction: instead of storing experience solely as natural-language
prompts, it materializes each learned strategy as an executable
pass in a standalone \python program, namely, the \emph{learned reducer}.
Since a reduction strategy is inherently tied to the syntax and
semantics of a specific language,
\eg, a transformation valid for \languageC is generally
inapplicable to other languages,
the learned reducer keeps a \emph{separate pool of
strategies} for each language and applies only the pool
matching the program under reduction.
This per-language organization does not compromise \proj's
language-agnostic nature, as every strategy pool is populated
automatically by the agent, without any hand-written,
language-specific engineering.

\begin{algorithm}[h!]
    \footnotesize
    \DontPrintSemicolon
    \SetKwInput{KwData}{Input}
    \SetKwInput{KwResult}{Output}
    \caption{
        The workflow of the learned reducer.
    }
    \label{alg:learned-reducer}

    \KwData{$\program$: the input program to be reduced.}
    \KwData{$\property$: the property to be preserved.}
    \KwData{$\transformations$: the accumulated strategy pool}
    \KwResult{$\programPrime$: the minimized program that satisfies $\property$.}

    \BlankLine
    $\programPrime \leftarrow \program$\;
    \Repeat{$\programPrime = P_{\mathit{prev}}$ \tcc*[f]{\small stop if no progress}}{
        $P_{\mathit{prev}} \leftarrow \programPrime$

        \ForEach{$\transformation = (\mathsf{match}_\transformation, \mathsf{rewrite}_\transformation) \in \transformations$}{
            \If(\tcc*[f]{\small pattern matching}){$\mathsf{match}_\transformation(\programPrime)$}{
                $\programCand \leftarrow \mathsf{rewrite}_\transformation(\programPrime)$ \tcc*[r]{\small rewrite}
                \If(){$\property(\programCand)$}{
                    $\programPrime \leftarrow \programCand$ \tcc*[r]{\small update $\programPrime$}\label{line:learned-reducer:update}
                }
            }
        }
    }
    \Return $\programPrime$

\end{algorithm}

\cref{alg:learned-reducer} shows the workflow of the learned reducer.
Given a program \program and a property \property,
the learned reducer repeatedly sweeps over the accumulated
strategy pool $\transformations$ to a fixpoint.
Each strategy
$\transformation = (\mathsf{match}_\transformation, \mathsf{rewrite}_\transformation)$
pairs a \emph{matching condition} with a \emph{rewrite}: whenever
$\mathsf{match}_\transformation$ holds on the current program,
the strategy applies $\mathsf{rewrite}_\transformation$ to produce
a candidate $\programCand$.
The candidate is accepted if it satisfies the property
(\ie, $\property(\programCand)$).
The process runs to a fixpoint~\cite{perses}
when no strategy can reduce the program further.
This design offers two key advantages over prompt-only memories:
(1) once implemented, a strategy can be invoked
without \llm queries, reducing cost and
token consumption; and
(2) each strategy has deterministic behavior
and remains guarded by the property checker,
eliminating the variability inherent
in \llm-based recall.

\mytightparagraph{Strategy Distillation.}
We now describe how the reflector agent distills each
successful reduction experience into a strategy in the learned reducer.
\cref{alg:reflector-agent} formalizes this process.
The reflector first reconstructs the completed case through a
series of \texttt{read\_*} tools (\cref{tab:tools}) that surface
the reducer's reduction attempts.
The reflector then examines each successful record
$r \in \mathcal{R}$
and decides whether the operation reflects a reusable reduction
pattern or is specific to the current program;
case-specific operations are skipped
(\cref{line:distillation:skip-case-specific}).

\begin{algorithm}[h!]
    \footnotesize
    \DontPrintSemicolon
    \SetKwInput{KwData}{Input}
    \SetKwInput{KwResult}{Output}
    \SetKwFunction{Generalize}{Generalize}
    \SetKwFunction{FindSimilar}{FindSimilar}
    \SetKwFunction{Covers}{Covers}
    \SetKwFunction{Strengthen}{Strengthen}
    \SetKwFunction{Patch}{Patch}
    \SetKwFunction{Verify}{Verify}
    \SetKwFunction{Quarantine}{Quarantine}
    \SetKwFunction{Commit}{Commit}
    \SetKwFunction{Refine}{Refine}
    \SetKwFunction{Discard}{Discard}

    \caption{
        The workflow of the reflector agent.
    }
    \label{alg:reflector-agent}

    \KwData{$\mathcal{R}$: successful reduction records from the reducer agent.}
    \KwData{$\transformations$: current strategy pool in the learned reducer.}
    \KwResult{$\transformations$: updated strategy pool.}

    \BlankLine

    \ForEach{record $r \in \mathcal{R}$}{
        \lIf{$r$ is case-specific}{\AlgContinue} \label{line:distillation:skip-case-specific}
        $\transformation_{\textit{new}} \leftarrow$ \Generalize{$r$}\tcc*[f]{\footnotesize generalization} \label{line:distillation:generalize}

        $\transformation_{\textit{sim}} \leftarrow$ \FindSimilar{$\transformation_{\textit{new}}, \transformations$}\tcc*[f]{\footnotesize find a similar strategy} \label{line:distillation:find-similar}

        \If{$\transformation_{\textit{sim}} \neq \mathrm{None}$}{
            $\transformation_{\textit{new}} \leftarrow$ \Patch{$\transformation_{\textit{sim}}, r$}\tcc*[f]{\footnotesize patch $\transformation_{\textit{sim}}$ to cover $r$} \label{line:distillation:strengthen}
        }

        \Repeat(\tcc*[f]{\footnotesize iterative strategy submission}){$\mathit{ok}$ or out of attempts}{
            $(\mathit{ok}, \mathit{feedback}) \leftarrow$ \Verify{$\transformation_{\textit{new}}$}\label{line:distillation:verify}

            \If{$\neg \mathit{ok}$}{$\transformation_{\textit{new}} \leftarrow$ \Refine{$\transformation_{\textit{new}}, \mathit{feedback}$}}\label{line:distillation:refine}
        }

        \If{$\mathit{ok}$}{
            $\transformations \leftarrow$ \Commit{$\transformations, \transformation_{\textit{new}}, \transformation_{\textit{sim}}$}\tcc*[f]{\footnotesize add $\transformation_{\textit{new}}$ or replace $\transformation_{\textit{sim}}$}\label{line:distillation:commit}
        }
    }
    \Return $\transformations$
\end{algorithm}

For each reusable record, the reflector generalizes the concrete
edit into an implementable strategy $\transformation_{\textit{new}}$
(\cref{line:distillation:generalize}): a \emph{matching condition}
that decides when the transformation applies, paired with a
\emph{rewrite rule} that transforms the matched fragment.
To keep generalization honest (\ie, to capture a \emph{pattern}
rather than fit the originating program), the reflector is
required to accompany each strategy with at least two tests,
at least one of which is drawn from a program other
than the original.
The reflector then checks the existing pool for a strategy that
already targets the same pattern
(\cref{line:distillation:find-similar}): if none exists, the
generalized strategy is kept as a new addition; if one exists
but does not yet cover the current record, the reflector patches
it to strengthen its coverage
(\cref{line:distillation:strengthen}).
Submitting a strategy is iterative: each
\texttt{propose\_action} is an \emph{attempt} that the harness
stages and runs through the verification pipeline (including
the tests above), returning either success or a structured
list of failures as feedback (\cref{line:distillation:verify}).
The reflector revises and
resubmits until the strategy passes, and only then lands it
into the learned reducer via \texttt{commit}
(\cref{line:distillation:commit}). Every distilled
strategy is thus not a mere natural-language reflection but an
executable, property-guarded reduction pass, reusable
in later sessions without \llm intervention.

\section{Evaluation}
\label{sec:evaluation}
This section presents the extensive evaluation of \proj to answer the
following research questions:
\begin{enumerate}[label=\textbf{RQ\arabic*}:, leftmargin=*]
    \item How does \proj perform in program reduction
    compared with the baselines?
    \item How does the learned reducer generalize to new cases?
    \item What is the contribution of each component in \proj?
\end{enumerate}

\subsection{Evaluation Setup}

We employ DeepSeek-V4-Flash~\cite{deepseek-v4-flash}
as the default \llm for \proj.
The maximum number of iterations is set to $I_{\max} = 3$
and the maximum number of attempts per iteration
is $A_{\max} = \maxTurnsFirstIteration,
\maxTurnsSecondIteration,
\maxTurnsThirdIteration$ for the first, second, and third
iterations, respectively.
The rationale for the decreasing schedule is that
the first iteration operates on a program that has not yet
been reduced by the agent, thus offering the most
reduction opportunities;
subsequent iterations work on progressively smaller programs
where fewer attempts suffice.
All experiments are conducted on an
Ubuntu 22.04 LTS server with
two Intel Xeon 6348 CPUs
and 512 GB of RAM.
For fairness, all experiments are
executed in a single-process, single-thread environment.

\mytightparagraph{Benchmark Suites.}
We collect \numBenchmarksAll benchmarks across three
programming languages.
\benchmarkC contains \numBenchmarksC \languageC programs,
each triggering a bug in \gcc or \llvm, including 20
benchmarks collected from prior program-reduction
work~\cite{perses,vulcan,lpr},
and 40 additional cases newly collected for this study.
\benchmarkRust comprises \numBenchmarksRust \languageRust programs
drawn from prior work~\cite{vulcan,lpr}, each triggering a bug
in \rustc.
\benchmarkJavaScript includes \numBenchmarksJavaScript
\languageJS programs also drawn from prior work~\cite{lpr},
each triggering a bug in \javascriptcore.
We randomly partition the 60 \languageC benchmarks into
two equal groups of 30:
\benchmarkCTrain is used in RQ1 for evaluating overall
performance, with the reflector agent enabled to accumulate
strategies during the process;
\benchmarkCTest is a held-out set reserved for RQ2
to evaluate the generalizability of the learned reducer
on programs the reducer has never seen.
The \benchmarkRust and \benchmarkJavaScript suites are
used entirely in RQ1, as their sizes are insufficient
for a meaningful train/test split.

\begin{table*}[!h]
    \centering
    \setlength{\tabcolsep}{9pt}
    \renewcommand{\arraystretch}{0.8}
    \caption{
        \small
        Evaluation results of \perses, \vulcan, \lpr, \creduce,
        and \proj.
        O(\#): \textbf{O}riginal program size.
        R(\#): \textbf{R}educed program size.
        T(s): execution \textbf{T}ime in seconds.
        C(\%): percentage \textbf{C}hange in size of \proj \wrt
        \vulcan, \lpr, and \creduce.
        Cells in green indicate the
        smallest reduction result among the techniques.
    }
    \label{tab:all}
    \resizebox{\textwidth}{!}{%
    \begin{tabular}{crrrrrrrrrrrrrr}
    \toprule
                                & \multicolumn{1}{c}{}                        & \multicolumn{2}{c}{\perses}                                                  & \multicolumn{2}{c}{\vulcan}                                                  & \multicolumn{2}{c}{\lpr}                                                     & \multicolumn{2}{c}{\creduce}                                                 & \multicolumn{2}{c}{\proj}                                                    & \multicolumn{3}{c}{C(\%) w.r.t.}                                                                                                      \\ \cmidrule(l){3-15} 
    \multirow{-2}{*}{Case} & \multicolumn{1}{c}{\multirow{-2}{*}{O(\#)}} & \multicolumn{1}{c}{\cellcolor[HTML]{EFEFEF}T(s)} & \multicolumn{1}{c}{R(\#)} & \multicolumn{1}{c}{\cellcolor[HTML]{EFEFEF}T(s)} & \multicolumn{1}{c}{R(\#)} & \multicolumn{1}{c}{\cellcolor[HTML]{EFEFEF}T(s)} & \multicolumn{1}{c}{R(\#)} & \multicolumn{1}{c}{\cellcolor[HTML]{EFEFEF}T(s)} & \multicolumn{1}{c}{R(\#)} & \multicolumn{1}{c}{\cellcolor[HTML]{EFEFEF}T(s)} & \multicolumn{1}{c}{R(\#)} & \multicolumn{1}{c}{\cellcolor[HTML]{DAE8FC}\vulcan} & \multicolumn{1}{c}{\cellcolor[HTML]{E3EEFE}\lpr} & \multicolumn{1}{c}{\cellcolor[HTML]{ECF4FF}\creduce} \\ \midrule
    C-1 & 23,146 & \cellcolor[HTML]{EFEFEF}723 & 203 & \cellcolor[HTML]{EFEFEF}2,479 & 185 & \cellcolor[HTML]{EFEFEF}1,541 & 209 & \cellcolor[HTML]{EFEFEF}4,727 & 183 & \cellcolor[HTML]{EFEFEF}1,687.7 & \cellcolor[HTML]{C6EFCE}66.3 & \cellcolor[HTML]{DAE8FC}-64.1\% & \cellcolor[HTML]{E3EEFE}-68.3\% & \cellcolor[HTML]{ECF4FF}-63.8\% \\
    C-2 & 6,133 & \cellcolor[HTML]{EFEFEF}40 & 51 & \cellcolor[HTML]{EFEFEF}72 & 38 & \cellcolor[HTML]{EFEFEF}699 & 48 & \cellcolor[HTML]{EFEFEF}306 & 46 & \cellcolor[HTML]{EFEFEF}644.0 & \cellcolor[HTML]{C6EFCE}27.0 & \cellcolor[HTML]{DAE8FC}-28.9\% & \cellcolor[HTML]{E3EEFE}-43.8\% & \cellcolor[HTML]{ECF4FF}-41.3\% \\
    C-3 & 36,735 & \cellcolor[HTML]{EFEFEF}3,629 & 174 & \cellcolor[HTML]{EFEFEF}32,869 & 147 & \cellcolor[HTML]{EFEFEF}6,158 & 156 & \cellcolor[HTML]{EFEFEF}20,376 & 85 & \cellcolor[HTML]{EFEFEF}4,651.1 & \cellcolor[HTML]{C6EFCE}56.3 & \cellcolor[HTML]{DAE8FC}-61.7\% & \cellcolor[HTML]{E3EEFE}-63.9\% & \cellcolor[HTML]{ECF4FF}-33.7\% \\
    C-4 & 25,709 & \cellcolor[HTML]{EFEFEF}1,191 & 318 & \cellcolor[HTML]{EFEFEF}9,293 & 206 & \cellcolor[HTML]{EFEFEF}7,812 & 143 & \cellcolor[HTML]{EFEFEF}5,510 & 143 & \cellcolor[HTML]{EFEFEF}2,454.5 & \cellcolor[HTML]{C6EFCE}60.7 & \cellcolor[HTML]{DAE8FC}-70.6\% & \cellcolor[HTML]{E3EEFE}-57.6\% & \cellcolor[HTML]{ECF4FF}-57.6\% \\
    C-5 & 148,931 & \cellcolor[HTML]{EFEFEF}2,932 & 239 & \cellcolor[HTML]{EFEFEF}13,326 & 203 & \cellcolor[HTML]{EFEFEF}5,913 & 151 & \cellcolor[HTML]{EFEFEF}6,948 & \cellcolor[HTML]{C6EFCE}78 & \cellcolor[HTML]{EFEFEF}4,055.4 & 148.3 & \cellcolor[HTML]{DAE8FC}-26.9\% & \cellcolor[HTML]{E3EEFE}-1.8\% & \cellcolor[HTML]{ECF4FF}+90.2\% \\
    C-6 & 11,748 & \cellcolor[HTML]{EFEFEF}893 & 288 & \cellcolor[HTML]{EFEFEF}19,066 & 205 & \cellcolor[HTML]{EFEFEF}12,367 & 190 & \cellcolor[HTML]{EFEFEF}6,664 & 160 & \cellcolor[HTML]{EFEFEF}2,102.9 & \cellcolor[HTML]{C6EFCE}45.0 & \cellcolor[HTML]{DAE8FC}-78.0\% & \cellcolor[HTML]{E3EEFE}-76.3\% & \cellcolor[HTML]{ECF4FF}-71.9\% \\
    C-7 & 173,840 & \cellcolor[HTML]{EFEFEF}1,914 & 117 & \cellcolor[HTML]{EFEFEF}4,597 & 79 & \cellcolor[HTML]{EFEFEF}4,507 & 82 & \cellcolor[HTML]{EFEFEF}3,095 & 68 & \cellcolor[HTML]{EFEFEF}3,105.1 & \cellcolor[HTML]{C6EFCE}50.7 & \cellcolor[HTML]{DAE8FC}-35.9\% & \cellcolor[HTML]{E3EEFE}-38.2\% & \cellcolor[HTML]{ECF4FF}-25.5\% \\
    C-8 & 23,655 & \cellcolor[HTML]{EFEFEF}708 & 237 & \cellcolor[HTML]{EFEFEF}2,556 & 224 & \cellcolor[HTML]{EFEFEF}1,788 & 243 & \cellcolor[HTML]{EFEFEF}4,768 & 193 & \cellcolor[HTML]{EFEFEF}1,938.7 & \cellcolor[HTML]{C6EFCE}99.7 & \cellcolor[HTML]{DAE8FC}-55.5\% & \cellcolor[HTML]{E3EEFE}-59.0\% & \cellcolor[HTML]{ECF4FF}-48.4\% \\
    C-9 & 24,203 & \cellcolor[HTML]{EFEFEF}833 & 259 & \cellcolor[HTML]{EFEFEF}5,446 & 170 & \cellcolor[HTML]{EFEFEF}2,890 & 238 & \cellcolor[HTML]{EFEFEF}5,527 & 168 & \cellcolor[HTML]{EFEFEF}2,051.2 & \cellcolor[HTML]{C6EFCE}94.0 & \cellcolor[HTML]{DAE8FC}-44.7\% & \cellcolor[HTML]{E3EEFE}-60.5\% & \cellcolor[HTML]{ECF4FF}-44.0\% \\
    C-10 & 174,538 & \cellcolor[HTML]{EFEFEF}8,302 & 180 & \cellcolor[HTML]{EFEFEF}23,402 & 88 & \cellcolor[HTML]{EFEFEF}13,477 & 103 & \cellcolor[HTML]{EFEFEF}13,675 & 50 & \cellcolor[HTML]{EFEFEF}9,201.8 & \cellcolor[HTML]{C6EFCE}42.0 & \cellcolor[HTML]{DAE8FC}-52.3\% & \cellcolor[HTML]{E3EEFE}-59.2\% & \cellcolor[HTML]{ECF4FF}-16.0\% \\
    C-11 & 209,577 & \cellcolor[HTML]{EFEFEF}2,081 & 120 & \cellcolor[HTML]{EFEFEF}3,587 & 56 & \cellcolor[HTML]{EFEFEF}3,832 & 46 & \cellcolor[HTML]{EFEFEF}2,596 & \cellcolor[HTML]{C6EFCE}43 & \cellcolor[HTML]{EFEFEF}2,993.2 & 53.0 & \cellcolor[HTML]{DAE8FC}-5.4\% & \cellcolor[HTML]{E3EEFE}+15.2\% & \cellcolor[HTML]{ECF4FF}+23.3\% \\
    C-12 & 20,029 & \cellcolor[HTML]{EFEFEF}2,573 & 479 & \cellcolor[HTML]{EFEFEF}31,733 & 302 & \cellcolor[HTML]{EFEFEF}6,804 & 368 & \cellcolor[HTML]{EFEFEF}8,775 & 185 & \cellcolor[HTML]{EFEFEF}3,915.9 & \cellcolor[HTML]{C6EFCE}141.0 & \cellcolor[HTML]{DAE8FC}-53.3\% & \cellcolor[HTML]{E3EEFE}-61.7\% & \cellcolor[HTML]{ECF4FF}-23.8\% \\
    C-13 & 25,943 & \cellcolor[HTML]{EFEFEF}1,065 & 332 & \cellcolor[HTML]{EFEFEF}4,150 & 218 & \cellcolor[HTML]{EFEFEF}6,546 & 133 & \cellcolor[HTML]{EFEFEF}5,241 & 143 & \cellcolor[HTML]{EFEFEF}2,071.2 & \cellcolor[HTML]{C6EFCE}101.0 & \cellcolor[HTML]{DAE8FC}-53.7\% & \cellcolor[HTML]{E3EEFE}-24.1\% & \cellcolor[HTML]{ECF4FF}-29.4\% \\
    C-14 & 85,359 & \cellcolor[HTML]{EFEFEF}1,055 & 150 & \cellcolor[HTML]{EFEFEF}2,012 & 103 & \cellcolor[HTML]{EFEFEF}2,360 & 68 & \cellcolor[HTML]{EFEFEF}4,575 & 65 & \cellcolor[HTML]{EFEFEF}2,131.7 & \cellcolor[HTML]{C6EFCE}59.3 & \cellcolor[HTML]{DAE8FC}-42.4\% & \cellcolor[HTML]{E3EEFE}-12.7\% & \cellcolor[HTML]{ECF4FF}-8.7\% \\
    C-15 & 27,386 & \cellcolor[HTML]{EFEFEF}2,598 & 238 & \cellcolor[HTML]{EFEFEF}20,739 & 221 & \cellcolor[HTML]{EFEFEF}7,054 & 76 & \cellcolor[HTML]{EFEFEF}9,853 & 150 & \cellcolor[HTML]{EFEFEF}3,808.3 & \cellcolor[HTML]{C6EFCE}70.0 & \cellcolor[HTML]{DAE8FC}-68.3\% & \cellcolor[HTML]{E3EEFE}-7.9\% & \cellcolor[HTML]{ECF4FF}-53.3\% \\
    C-16 & 25,437 & \cellcolor[HTML]{EFEFEF}981 & 332 & \cellcolor[HTML]{EFEFEF}5,274 & 218 & \cellcolor[HTML]{EFEFEF}5,912 & 167 & \cellcolor[HTML]{EFEFEF}4,086 & 143 & \cellcolor[HTML]{EFEFEF}2,000.6 & \cellcolor[HTML]{C6EFCE}106.0 & \cellcolor[HTML]{DAE8FC}-51.4\% & \cellcolor[HTML]{E3EEFE}-36.5\% & \cellcolor[HTML]{ECF4FF}-25.9\% \\
    C-17 & 30,196 & \cellcolor[HTML]{EFEFEF}839 & 98 & \cellcolor[HTML]{EFEFEF}1,811 & 91 & \cellcolor[HTML]{EFEFEF}1,761 & 75 & \cellcolor[HTML]{EFEFEF}1,536 & 73 & \cellcolor[HTML]{EFEFEF}1,752.5 & \cellcolor[HTML]{C6EFCE}61.0 & \cellcolor[HTML]{DAE8FC}-33.0\% & \cellcolor[HTML]{E3EEFE}-18.7\% & \cellcolor[HTML]{ECF4FF}-16.4\% \\
    C-18 & 38,647 & \cellcolor[HTML]{EFEFEF}2,668 & 464 & \cellcolor[HTML]{EFEFEF}24,420 & 303 & \cellcolor[HTML]{EFEFEF}5,447 & 98 & \cellcolor[HTML]{EFEFEF}6,716 & 118 & \cellcolor[HTML]{EFEFEF}3,771.2 & \cellcolor[HTML]{C6EFCE}78.0 & \cellcolor[HTML]{DAE8FC}-74.3\% & \cellcolor[HTML]{E3EEFE}-20.4\% & \cellcolor[HTML]{ECF4FF}-33.9\% \\
    C-19 & 19,585 & \cellcolor[HTML]{EFEFEF}2,141 & 457 & \cellcolor[HTML]{EFEFEF}8,310 & 364 & \cellcolor[HTML]{EFEFEF}11,420 & 395 & \cellcolor[HTML]{EFEFEF}7,999 & 202 & \cellcolor[HTML]{EFEFEF}3,355.6 & \cellcolor[HTML]{C6EFCE}185.0 & \cellcolor[HTML]{DAE8FC}-49.2\% & \cellcolor[HTML]{E3EEFE}-53.2\% & \cellcolor[HTML]{ECF4FF}-8.4\% \\
    C-20 & 29,625 & \cellcolor[HTML]{EFEFEF}1,482 & 289 & \cellcolor[HTML]{EFEFEF}17,262 & 180 & \cellcolor[HTML]{EFEFEF}7,344 & 274 & \cellcolor[HTML]{EFEFEF}9,247 & 178 & \cellcolor[HTML]{EFEFEF}2,931.8 & \cellcolor[HTML]{C6EFCE}46.0 & \cellcolor[HTML]{DAE8FC}-74.4\% & \cellcolor[HTML]{E3EEFE}-83.2\% & \cellcolor[HTML]{ECF4FF}-74.2\% \\
    C-21 & 43,942 & \cellcolor[HTML]{EFEFEF}1,159 & 153 & \cellcolor[HTML]{EFEFEF}8,904 & 84 & \cellcolor[HTML]{EFEFEF}2,910 & 69 & \cellcolor[HTML]{EFEFEF}3,890 & 53 & \cellcolor[HTML]{EFEFEF}1,898.2 & \cellcolor[HTML]{C6EFCE}46.0 & \cellcolor[HTML]{DAE8FC}-45.2\% & \cellcolor[HTML]{E3EEFE}-33.3\% & \cellcolor[HTML]{ECF4FF}-13.2\% \\
    C-22 & 22,274 & \cellcolor[HTML]{EFEFEF}499 & 244 & \cellcolor[HTML]{EFEFEF}2,829 & 223 & \cellcolor[HTML]{EFEFEF}2,146 & 203 & \cellcolor[HTML]{EFEFEF}4,223 & 197 & \cellcolor[HTML]{EFEFEF}1,598.5 & \cellcolor[HTML]{C6EFCE}66.0 & \cellcolor[HTML]{DAE8FC}-70.4\% & \cellcolor[HTML]{E3EEFE}-67.5\% & \cellcolor[HTML]{ECF4FF}-66.5\% \\
    C-23 & 78,960 & \cellcolor[HTML]{EFEFEF}640 & 239 & \cellcolor[HTML]{EFEFEF}2,410 & 104 & \cellcolor[HTML]{EFEFEF}2,442 & 99 & \cellcolor[HTML]{EFEFEF}1,665 & 90 & \cellcolor[HTML]{EFEFEF}1,650.6 & \cellcolor[HTML]{C6EFCE}67.7 & \cellcolor[HTML]{DAE8FC}-34.9\% & \cellcolor[HTML]{E3EEFE}-31.6\% & \cellcolor[HTML]{ECF4FF}-24.8\% \\
    C-24 & 6,592 & \cellcolor[HTML]{EFEFEF}1,689 & 413 & \cellcolor[HTML]{EFEFEF}5,416 & 377 & \cellcolor[HTML]{EFEFEF}9,593 & 385 & \cellcolor[HTML]{EFEFEF}9,541 & 235 & \cellcolor[HTML]{EFEFEF}3,196.0 & \cellcolor[HTML]{C6EFCE}152.7 & \cellcolor[HTML]{DAE8FC}-59.5\% & \cellcolor[HTML]{E3EEFE}-60.3\% & \cellcolor[HTML]{ECF4FF}-35.0\% \\
    C-25 & 20,287 & \cellcolor[HTML]{EFEFEF}753 & 203 & \cellcolor[HTML]{EFEFEF}2,293 & 185 & \cellcolor[HTML]{EFEFEF}2,383 & 185 & \cellcolor[HTML]{EFEFEF}4,514 & 183 & \cellcolor[HTML]{EFEFEF}1,926.8 & \cellcolor[HTML]{C6EFCE}64.7 & \cellcolor[HTML]{DAE8FC}-65.0\% & \cellcolor[HTML]{E3EEFE}-65.0\% & \cellcolor[HTML]{ECF4FF}-64.7\% \\
    C-26 & 20,729 & \cellcolor[HTML]{EFEFEF}503 & 244 & \cellcolor[HTML]{EFEFEF}1,559 & 223 & \cellcolor[HTML]{EFEFEF}4,684 & 195 & \cellcolor[HTML]{EFEFEF}4,255 & 197 & \cellcolor[HTML]{EFEFEF}1,514.8 & \cellcolor[HTML]{C6EFCE}65.7 & \cellcolor[HTML]{DAE8FC}-70.6\% & \cellcolor[HTML]{E3EEFE}-66.3\% & \cellcolor[HTML]{ECF4FF}-66.7\% \\
    C-27 & 15,770 & \cellcolor[HTML]{EFEFEF}660 & 373 & \cellcolor[HTML]{EFEFEF}3,145 & 345 & \cellcolor[HTML]{EFEFEF}4,110 & 375 & \cellcolor[HTML]{EFEFEF}5,582 & \cellcolor[HTML]{C6EFCE}173 & \cellcolor[HTML]{EFEFEF}1,979.3 & 181.0 & \cellcolor[HTML]{DAE8FC}-47.5\% & \cellcolor[HTML]{E3EEFE}-51.7\% & \cellcolor[HTML]{ECF4FF}+4.6\% \\
    C-28 & 18,446 & \cellcolor[HTML]{EFEFEF}2,354 & 342 & \cellcolor[HTML]{EFEFEF}16,809 & 329 & \cellcolor[HTML]{EFEFEF}12,930 & 315 & \cellcolor[HTML]{EFEFEF}13,220 & 127 & \cellcolor[HTML]{EFEFEF}3,657.9 & \cellcolor[HTML]{C6EFCE}65.7 & \cellcolor[HTML]{DAE8FC}-80.0\% & \cellcolor[HTML]{E3EEFE}-79.2\% & \cellcolor[HTML]{ECF4FF}-48.3\% \\
    C-29 & 7,252 & \cellcolor[HTML]{EFEFEF}481 & 222 & \cellcolor[HTML]{EFEFEF}4,932 & 218 & \cellcolor[HTML]{EFEFEF}6,462 & 186 & \cellcolor[HTML]{EFEFEF}2,462 & 143 & \cellcolor[HTML]{EFEFEF}1,526.1 & \cellcolor[HTML]{C6EFCE}100.3 & \cellcolor[HTML]{DAE8FC}-54.0\% & \cellcolor[HTML]{E3EEFE}-46.1\% & \cellcolor[HTML]{ECF4FF}-29.8\% \\
    C-30 & 24,492 & \cellcolor[HTML]{EFEFEF}726 & 168 & \cellcolor[HTML]{EFEFEF}1,989 & 153 & \cellcolor[HTML]{EFEFEF}1,973 & 145 & \cellcolor[HTML]{EFEFEF}6,246 & 168 & \cellcolor[HTML]{EFEFEF}1,721.1 & \cellcolor[HTML]{C6EFCE}65.7 & \cellcolor[HTML]{DAE8FC}-57.1\% & \cellcolor[HTML]{E3EEFE}-54.7\% & \cellcolor[HTML]{ECF4FF}-60.9\% \\
    \midrule
    Mean & 47,306 & \cellcolor[HTML]{EFEFEF}1,604 & 254 & \cellcolor[HTML]{EFEFEF}9,423 & 195 & \cellcolor[HTML]{EFEFEF}5,509 & 181 & \cellcolor[HTML]{EFEFEF}6,261 & 135 & \cellcolor[HTML]{EFEFEF}2,709.8 & \cellcolor[HTML]{C6EFCE}82.2 & \cellcolor[HTML]{DAE8FC}-57.8\% & \cellcolor[HTML]{E3EEFE}-54.5\% & \cellcolor[HTML]{ECF4FF}-39.0\% \\
    \midrule
    Rust-1 & 190 & \cellcolor[HTML]{EFEFEF}832 & 114 & \cellcolor[HTML]{EFEFEF}9,235 & 101 & \cellcolor[HTML]{EFEFEF}2,638 & 97 & \cellcolor[HTML]{EFEFEF}819 & 109 & \cellcolor[HTML]{EFEFEF}1,920.4 & \cellcolor[HTML]{C6EFCE}74.7 & \cellcolor[HTML]{DAE8FC}-26.1\% & \cellcolor[HTML]{E3EEFE}-23.0\% & \cellcolor[HTML]{ECF4FF}-31.5\% \\
    Rust-2 & 192 & \cellcolor[HTML]{EFEFEF}45 & 166 & \cellcolor[HTML]{EFEFEF}729 & 157 & \cellcolor[HTML]{EFEFEF}2,546 & 115 & \cellcolor[HTML]{EFEFEF}744 & 161 & \cellcolor[HTML]{EFEFEF}1,013.4 & \cellcolor[HTML]{C6EFCE}66.7 & \cellcolor[HTML]{DAE8FC}-57.5\% & \cellcolor[HTML]{E3EEFE}-42.0\% & \cellcolor[HTML]{ECF4FF}-58.6\% \\
    Rust-3 & 513 & \cellcolor[HTML]{EFEFEF}1,485 & 174 & \cellcolor[HTML]{EFEFEF}2,583 & 82 & \cellcolor[HTML]{EFEFEF}2,492 & 74 & \cellcolor[HTML]{EFEFEF}18,803 & 101 & \cellcolor[HTML]{EFEFEF}2,158.4 & \cellcolor[HTML]{C6EFCE}62.7 & \cellcolor[HTML]{DAE8FC}-23.6\% & \cellcolor[HTML]{E3EEFE}-15.3\% & \cellcolor[HTML]{ECF4FF}-38.0\% \\
    Rust-4 & 263 & \cellcolor[HTML]{EFEFEF}10 & 56 & \cellcolor[HTML]{EFEFEF}104 & \cellcolor[HTML]{E2F4E5}49 & \cellcolor[HTML]{EFEFEF}159 & 56 & \cellcolor[HTML]{EFEFEF}204 & \cellcolor[HTML]{E2F4E5}49 & \cellcolor[HTML]{EFEFEF}712.1 & 56.3 & \cellcolor[HTML]{DAE8FC}+15.0\% & \cellcolor[HTML]{E3EEFE}+0.6\% & \cellcolor[HTML]{ECF4FF}+15.0\% \\
    Rust-5 & 81 & \cellcolor[HTML]{EFEFEF}3 & \cellcolor[HTML]{E2F4E5}13 & \cellcolor[HTML]{EFEFEF}14 & \cellcolor[HTML]{E2F4E5}13 & \cellcolor[HTML]{EFEFEF}45 & \cellcolor[HTML]{E2F4E5}13 & \cellcolor[HTML]{EFEFEF}30 & \cellcolor[HTML]{E2F4E5}13 & \cellcolor[HTML]{EFEFEF}291.3 & \cellcolor[HTML]{E2F4E5}13.0 & \cellcolor[HTML]{DAE8FC}+0.0\% & \cellcolor[HTML]{E3EEFE}+0.0\% & \cellcolor[HTML]{ECF4FF}+0.0\% \\
    Rust-6 & 63 & \cellcolor[HTML]{EFEFEF}9 & 34 & \cellcolor[HTML]{EFEFEF}57 & \cellcolor[HTML]{C6EFCE}21 & \cellcolor[HTML]{EFEFEF}565 & 34 & \cellcolor[HTML]{EFEFEF}99 & 23 & \cellcolor[HTML]{EFEFEF}508.1 & 23.0 & \cellcolor[HTML]{DAE8FC}+9.5\% & \cellcolor[HTML]{E3EEFE}-32.4\% & \cellcolor[HTML]{ECF4FF}+0.0\% \\
    Rust-7 & 644 & \cellcolor[HTML]{EFEFEF}4,972 & 382 & \cellcolor[HTML]{EFEFEF}18,773 & 338 & \cellcolor[HTML]{EFEFEF}11,208 & 379 & \cellcolor[HTML]{EFEFEF}11,133 & 537 & \cellcolor[HTML]{EFEFEF}6,451.6 & \cellcolor[HTML]{C6EFCE}188.7 & \cellcolor[HTML]{DAE8FC}-44.2\% & \cellcolor[HTML]{E3EEFE}-50.2\% & \cellcolor[HTML]{ECF4FF}-64.9\% \\
    Rust-8 & 182 & \cellcolor[HTML]{EFEFEF}12 & 102 & \cellcolor[HTML]{EFEFEF}129 & 102 & \cellcolor[HTML]{EFEFEF}1,190 & 102 & \cellcolor[HTML]{EFEFEF}273 & \cellcolor[HTML]{C6EFCE}75 & \cellcolor[HTML]{EFEFEF}530.5 & 80.3 & \cellcolor[HTML]{DAE8FC}-21.2\% & \cellcolor[HTML]{E3EEFE}-21.2\% & \cellcolor[HTML]{ECF4FF}+7.1\% \\
    Rust-9 & 936 & \cellcolor[HTML]{EFEFEF}5,395 & 728 & \cellcolor[HTML]{EFEFEF}49,846 & 713 & \cellcolor[HTML]{EFEFEF}8,431 & 734 & \cellcolor[HTML]{EFEFEF}8,629 & 649 & \cellcolor[HTML]{EFEFEF}7,340.4 & \cellcolor[HTML]{C6EFCE}480.7 & \cellcolor[HTML]{DAE8FC}-32.6\% & \cellcolor[HTML]{E3EEFE}-34.5\% & \cellcolor[HTML]{ECF4FF}-25.9\% \\
    Rust-10 & 65 & \cellcolor[HTML]{EFEFEF}3 & 28 & \cellcolor[HTML]{EFEFEF}34 & 26 & \cellcolor[HTML]{EFEFEF}123 & 29 & \cellcolor[HTML]{EFEFEF}125 & 34 & \cellcolor[HTML]{EFEFEF}613.0 & \cellcolor[HTML]{C6EFCE}25.3 & \cellcolor[HTML]{DAE8FC}-2.6\% & \cellcolor[HTML]{E3EEFE}-12.6\% & \cellcolor[HTML]{ECF4FF}-25.5\% \\
    Rust-11 & 132 & \cellcolor[HTML]{EFEFEF}13 & 74 & \cellcolor[HTML]{EFEFEF}177 & 62 & \cellcolor[HTML]{EFEFEF}544 & 67 & \cellcolor[HTML]{EFEFEF}410 & 70 & \cellcolor[HTML]{EFEFEF}810.1 & \cellcolor[HTML]{C6EFCE}43.0 & \cellcolor[HTML]{DAE8FC}-30.6\% & \cellcolor[HTML]{E3EEFE}-35.8\% & \cellcolor[HTML]{ECF4FF}-38.6\% \\
    Rust-12 & 347 & \cellcolor[HTML]{EFEFEF}225 & 263 & \cellcolor[HTML]{EFEFEF}1,400 & 247 & \cellcolor[HTML]{EFEFEF}1,920 & 263 & \cellcolor[HTML]{EFEFEF}2,615 & 264 & \cellcolor[HTML]{EFEFEF}1,249.0 & \cellcolor[HTML]{C6EFCE}56.7 & \cellcolor[HTML]{DAE8FC}-77.1\% & \cellcolor[HTML]{E3EEFE}-78.5\% & \cellcolor[HTML]{ECF4FF}-78.5\% \\
    Rust-13 & 173 & \cellcolor[HTML]{EFEFEF}6 & 39 & \cellcolor[HTML]{EFEFEF}87 & 39 & \cellcolor[HTML]{EFEFEF}411 & 40 & \cellcolor[HTML]{EFEFEF}184 & 39 & \cellcolor[HTML]{EFEFEF}690.4 & \cellcolor[HTML]{C6EFCE}28.3 & \cellcolor[HTML]{DAE8FC}-27.4\% & \cellcolor[HTML]{E3EEFE}-29.2\% & \cellcolor[HTML]{ECF4FF}-27.4\% \\
    Rust-14 & 556 & \cellcolor[HTML]{EFEFEF}3,730 & 458 & \cellcolor[HTML]{EFEFEF}32,524 & 442 & \cellcolor[HTML]{EFEFEF}5,956 & 463 & \cellcolor[HTML]{EFEFEF}8,565 & 447 & \cellcolor[HTML]{EFEFEF}5,440.0 & \cellcolor[HTML]{C6EFCE}259.7 & \cellcolor[HTML]{DAE8FC}-41.3\% & \cellcolor[HTML]{E3EEFE}-43.9\% & \cellcolor[HTML]{ECF4FF}-41.9\% \\
    Rust-15 & 801 & \cellcolor[HTML]{EFEFEF}638 & 467 & \cellcolor[HTML]{EFEFEF}7,271 & 284 & \cellcolor[HTML]{EFEFEF}2,639 & 469 & \cellcolor[HTML]{EFEFEF}6,999 & 471 & \cellcolor[HTML]{EFEFEF}2,153.1 & \cellcolor[HTML]{C6EFCE}103.0 & \cellcolor[HTML]{DAE8FC}-63.7\% & \cellcolor[HTML]{E3EEFE}-78.0\% & \cellcolor[HTML]{ECF4FF}-78.1\% \\
    Rust-16 & 28 & \cellcolor[HTML]{EFEFEF}3 & \cellcolor[HTML]{E2F4E5}26 & \cellcolor[HTML]{EFEFEF}53 & \cellcolor[HTML]{E2F4E5}26 & \cellcolor[HTML]{EFEFEF}378 & \cellcolor[HTML]{E2F4E5}26 & \cellcolor[HTML]{EFEFEF}105 & \cellcolor[HTML]{E2F4E5}26 & \cellcolor[HTML]{EFEFEF}606.1 & \cellcolor[HTML]{E2F4E5}26.0 & \cellcolor[HTML]{DAE8FC}+0.0\% & \cellcolor[HTML]{E3EEFE}+0.0\% & \cellcolor[HTML]{ECF4FF}+0.0\% \\
    Rust-17 & 448 & \cellcolor[HTML]{EFEFEF}6,859 & 299 & \cellcolor[HTML]{EFEFEF}78,553 & 277 & \cellcolor[HTML]{EFEFEF}12,125 & 263 & \cellcolor[HTML]{EFEFEF}60,254 & \cellcolor[HTML]{C6EFCE}160 & \cellcolor[HTML]{EFEFEF}8,544.5 & 197.0 & \cellcolor[HTML]{DAE8FC}-28.9\% & \cellcolor[HTML]{E3EEFE}-25.1\% & \cellcolor[HTML]{ECF4FF}+23.1\% \\
    Rust-18 & 957 & \cellcolor[HTML]{EFEFEF}7 & 17 & \cellcolor[HTML]{EFEFEF}29 & \cellcolor[HTML]{C6EFCE}9 & \cellcolor[HTML]{EFEFEF}167 & 17 & \cellcolor[HTML]{EFEFEF}92 & 12 & \cellcolor[HTML]{EFEFEF}310.4 & 9.3 & \cellcolor[HTML]{DAE8FC}+3.7\% & \cellcolor[HTML]{E3EEFE}-45.1\% & \cellcolor[HTML]{ECF4FF}-22.2\% \\
    Rust-19 & 866 & \cellcolor[HTML]{EFEFEF}5,924 & 736 & \cellcolor[HTML]{EFEFEF}101,313 & 635 & \cellcolor[HTML]{EFEFEF}8,524 & 741 & \cellcolor[HTML]{EFEFEF}7,508 & 726 & \cellcolor[HTML]{EFEFEF}7,754.5 & \cellcolor[HTML]{C6EFCE}504.3 & \cellcolor[HTML]{DAE8FC}-20.6\% & \cellcolor[HTML]{E3EEFE}-31.9\% & \cellcolor[HTML]{ECF4FF}-30.5\% \\
    Rust-20 & 121 & \cellcolor[HTML]{EFEFEF}15 & 72 & \cellcolor[HTML]{EFEFEF}243 & 56 & \cellcolor[HTML]{EFEFEF}1,496 & 58 & \cellcolor[HTML]{EFEFEF}541 & \cellcolor[HTML]{C6EFCE}51 & \cellcolor[HTML]{EFEFEF}822.9 & 55.0 & \cellcolor[HTML]{DAE8FC}-1.8\% & \cellcolor[HTML]{E3EEFE}-5.2\% & \cellcolor[HTML]{ECF4FF}+7.8\% \\
    \midrule
    Mean & 378 & \cellcolor[HTML]{EFEFEF}1,509 & 212 & \cellcolor[HTML]{EFEFEF}15,158 & 184 & \cellcolor[HTML]{EFEFEF}3,178 & 202 & \cellcolor[HTML]{EFEFEF}6,407 & 201 & \cellcolor[HTML]{EFEFEF}2,496.0 & \cellcolor[HTML]{C6EFCE}117.7 & \cellcolor[HTML]{DAE8FC}-36.0\% & \cellcolor[HTML]{E3EEFE}-41.7\% & \cellcolor[HTML]{ECF4FF}-41.4\% \\
    \midrule
    JS-1 & 63 & \cellcolor[HTML]{EFEFEF}193 & 56 & \cellcolor[HTML]{EFEFEF}2,791 & 41 & \cellcolor[HTML]{EFEFEF}1,801 & 45 & \cellcolor[HTML]{EFEFEF}1,454 & 43 & \cellcolor[HTML]{EFEFEF}996.9 & \cellcolor[HTML]{C6EFCE}24.0 & \cellcolor[HTML]{DAE8FC}-41.5\% & \cellcolor[HTML]{E3EEFE}-46.7\% & \cellcolor[HTML]{ECF4FF}-44.2\% \\
    JS-2 & 185 & \cellcolor[HTML]{EFEFEF}42 & 65 & \cellcolor[HTML]{EFEFEF}3,215 & 35 & \cellcolor[HTML]{EFEFEF}2,064 & 48 & \cellcolor[HTML]{EFEFEF}410 & 47 & \cellcolor[HTML]{EFEFEF}720.4 & \cellcolor[HTML]{C6EFCE}23.3 & \cellcolor[HTML]{DAE8FC}-33.3\% & \cellcolor[HTML]{E3EEFE}-51.4\% & \cellcolor[HTML]{ECF4FF}-50.4\% \\
    JS-3 & 244 & \cellcolor[HTML]{EFEFEF}175 & 52 & \cellcolor[HTML]{EFEFEF}5,631 & 41 & \cellcolor[HTML]{EFEFEF}2,301 & 30 & \cellcolor[HTML]{EFEFEF}1,293 & 41 & \cellcolor[HTML]{EFEFEF}1,002.4 & \cellcolor[HTML]{C6EFCE}21.7 & \cellcolor[HTML]{DAE8FC}-47.2\% & \cellcolor[HTML]{E3EEFE}-27.8\% & \cellcolor[HTML]{ECF4FF}-47.2\% \\
    JS-4 & 125 & \cellcolor[HTML]{EFEFEF}165 & 57 & \cellcolor[HTML]{EFEFEF}5,098 & 41 & \cellcolor[HTML]{EFEFEF}1,041 & 46 & \cellcolor[HTML]{EFEFEF}1,110 & 47 & \cellcolor[HTML]{EFEFEF}1,172.7 & \cellcolor[HTML]{C6EFCE}21.7 & \cellcolor[HTML]{DAE8FC}-47.2\% & \cellcolor[HTML]{E3EEFE}-52.9\% & \cellcolor[HTML]{ECF4FF}-53.9\% \\
    JS-5 & 178 & \cellcolor[HTML]{EFEFEF}31 & 66 & \cellcolor[HTML]{EFEFEF}1,010 & 38 & \cellcolor[HTML]{EFEFEF}2,290 & 43 & \cellcolor[HTML]{EFEFEF}482 & 52 & \cellcolor[HTML]{EFEFEF}676.6 & \cellcolor[HTML]{C6EFCE}27.0 & \cellcolor[HTML]{DAE8FC}-28.9\% & \cellcolor[HTML]{E3EEFE}-37.2\% & \cellcolor[HTML]{ECF4FF}-48.1\% \\
    JS-6 & 112 & \cellcolor[HTML]{EFEFEF}276 & 51 & \cellcolor[HTML]{EFEFEF}2,726 & 41 & \cellcolor[HTML]{EFEFEF}1,249 & 41 & \cellcolor[HTML]{EFEFEF}1,860 & 40 & \cellcolor[HTML]{EFEFEF}1,118.4 & \cellcolor[HTML]{C6EFCE}26.3 & \cellcolor[HTML]{DAE8FC}-35.8\% & \cellcolor[HTML]{E3EEFE}-35.8\% & \cellcolor[HTML]{ECF4FF}-34.2\% \\
    JS-7 & 152 & \cellcolor[HTML]{EFEFEF}255 & 57 & \cellcolor[HTML]{EFEFEF}3,183 & 30 & \cellcolor[HTML]{EFEFEF}1,116 & 34 & \cellcolor[HTML]{EFEFEF}1,973 & 45 & \cellcolor[HTML]{EFEFEF}717.9 & \cellcolor[HTML]{C6EFCE}19.7 & \cellcolor[HTML]{DAE8FC}-34.4\% & \cellcolor[HTML]{E3EEFE}-42.2\% & \cellcolor[HTML]{ECF4FF}-56.3\% \\
    JS-8 & 87 & \cellcolor[HTML]{EFEFEF}149 & 50 & \cellcolor[HTML]{EFEFEF}2,149 & 30 & \cellcolor[HTML]{EFEFEF}696 & 50 & \cellcolor[HTML]{EFEFEF}1,295 & 32 & \cellcolor[HTML]{EFEFEF}748.7 & \cellcolor[HTML]{C6EFCE}20.0 & \cellcolor[HTML]{DAE8FC}-33.3\% & \cellcolor[HTML]{E3EEFE}-60.0\% & \cellcolor[HTML]{ECF4FF}-37.5\% \\
    JS-9 & 121 & \cellcolor[HTML]{EFEFEF}97 & 55 & \cellcolor[HTML]{EFEFEF}3,303 & 47 & \cellcolor[HTML]{EFEFEF}2,279 & 35 & \cellcolor[HTML]{EFEFEF}637 & 45 & \cellcolor[HTML]{EFEFEF}1,139.0 & \cellcolor[HTML]{C6EFCE}28.0 & \cellcolor[HTML]{DAE8FC}-40.4\% & \cellcolor[HTML]{E3EEFE}-20.0\% & \cellcolor[HTML]{ECF4FF}-37.8\% \\
    JS-10 & 144 & \cellcolor[HTML]{EFEFEF}35 & 46 & \cellcolor[HTML]{EFEFEF}653 & 38 & \cellcolor[HTML]{EFEFEF}890 & 34 & \cellcolor[HTML]{EFEFEF}5,436 & 38 & \cellcolor[HTML]{EFEFEF}953.5 & \cellcolor[HTML]{C6EFCE}21.7 & \cellcolor[HTML]{DAE8FC}-43.0\% & \cellcolor[HTML]{E3EEFE}-36.3\% & \cellcolor[HTML]{ECF4FF}-43.0\% \\
    \midrule
    Mean & 141 & \cellcolor[HTML]{EFEFEF}142 & 56 & \cellcolor[HTML]{EFEFEF}2,976 & 38 & \cellcolor[HTML]{EFEFEF}1,573 & 41 & \cellcolor[HTML]{EFEFEF}1,595 & 43 & \cellcolor[HTML]{EFEFEF}924.7 & \cellcolor[HTML]{C6EFCE}23.3 & \cellcolor[HTML]{DAE8FC}-38.9\% & \cellcolor[HTML]{E3EEFE}-42.5\% & \cellcolor[HTML]{ECF4FF}-45.7\% \\
    \bottomrule
    \end{tabular}%
    }
    \end{table*}

\mytightparagraph{Baselines.}
We evaluate \proj against four baselines:

\begin{itemize}[leftmargin=*, topsep=-0.5pt, itemsep=0pt, partopsep=0pt]
    \item \emph{\perses}
    is a language-agnostic syntax-guided program
    reducer~\cite{perses}. It excels at efficiency,
    and reducers including \vulcan and \lpr are
    built on top of \perses.

    \item \emph{\vulcan}~\cite{vulcan}
    is a language-agnostic program reducer based on \perses.
    It improves the effectiveness of \perses
    with aggressive language-agnostic transformations.

    \item \emph{\lpr}~\cite{lpr}
    is an \llm-assisted reducer built on top of \perses.
    It employs an \llm to perform a series of predefined
    code transformations. We equip \lpr with the same model
    as \proj, \ie, \deepseekVfourflash,
    for a fair comparison.

    \item \emph{\creduce}~\cite{creduce}
    is the state-of-the-art program reducer for C programs.
    \creduce also contains certain
    language-agnostic components, enabling it to
    handle programs in other programming languages.

\end{itemize}

\mytightparagraph{Metrics.}
We evaluate \proj using the following metrics:

\begin{itemize}[leftmargin=*, topsep=-0.5pt, itemsep=0pt, partopsep=0pt]
    \item \emph{Effectiveness}:
    we use \textbf{R(\#)}, the final number of lexical
    tokens in the reduced program to measure the effectiveness;
    the smaller the R(\#) is,
    the more effective the reducer is.

    \item \emph{Efficiency}:
    We measure the efficiency of a reducer by
    \textbf{T(s)}, the duration of the reduction
    process in seconds. The smaller the T(s) is,
    the more efficient the reducer is.

    \item \emph{Cost}:
    For LLM-assisted reducers, \ie, \proj and \lpr,
    we measure their cost by \textbf{C(USD)}, the total
    cost of the LLM API calls during the reduction process.

\end{itemize}

\subsection{RQ1: Overall Performance}

We evaluate the
effectiveness, efficiency, and cost
of \proj.
We use \benchmarkCTrain (30 \languageC programs),
\benchmarkRust (20), and \benchmarkJavaScript (10).

\mytightparagraph{Methodology.}
To faithfully simulate how \proj would be used in practice,
we process benchmarks \emph{sequentially} within each
language suite with the reflector agent enabled.
Concretely, we randomly order the benchmarks,
then reduce them one by one;
after each case completes, the reflector agent distills
successful reduction experiences into the learned reducer
before the next case begins.
As a result, each subsequent case benefits from the
strategies accumulated by all preceding cases,
mirroring a typical usage scenario of \proj
in real-world settings.
To mitigate the effect of ordering randomness,
we repeat the entire experiment
\emph{three times with independently shuffled orderings}
and report the mean results.

\mytightparagraph{Effectiveness.}
\cref{tab:all} shows the evaluation results,
with the reduction results listed in R(\#) columns.
In general,
\proj is the most effective reducer by a wide and consistent
margin: it produces the smallest program on every benchmark
suite and against every baseline.
On average, \proj reduces the programs to
\astraAverageSizeC, \astraAverageSizeRust, and
\astraAverageSizeJavaScript tokens on the
\languageC, \languageRust, and \languageJS benchmark suites,
respectively. These results are
\astraImprovementOverBestBaselineC,
\astraImprovementOverBestBaselineRust, and
\astraImprovementOverBestBaselineJavaScript smaller
than the best baseline on the corresponding language suite,
respectively.
These are large margins for a mature task in which existing
reducers are already highly optimized: \proj removes a third or
more of the tokens that the best prior reducer leaves behind.
Notably, to the best of our knowledge, \proj is the first
reducer to outperform \creduce on \languageC programs.
As a reducer carefully specialized for \languageC,
\creduce has long been
regarded as the de facto standard for \languageC program reduction.
Surpassing \creduce therefore provides strong evidence that
\proj's case-specific, semantic-aware reduction can identify
opportunities beyond those captured by even highly engineered
language-specific reducers.
Moreover, \proj also decisively outperforms the other
\llm-assisted reducer \lpr, directly demonstrating the
advantage of agentic, per-program reasoning
over a fixed repertoire of predefined transformations.

\mytightparagraph{Efficiency.}
The T(s) columns in \cref{tab:all} list
the processing time of each approach.
\perses is the fastest reducer, which is expected:
\vulcan, \lpr, and \proj all invoke \perses as their
initial reduction stage, and therefore cannot be faster
than \perses alone.
As a highly efficient syntax-guided reducer,
\perses is also generally faster than \creduce.
Excluding \perses, \proj achieves the best efficiency among
all approaches, reducing average running time by
\astraImprovementOverVulcanTime,
\astraImprovementOverLPRTime, and
\astraImprovementOverCreduceTime compared with
\vulcan, \lpr, and \creduce, respectively.
More importantly, on complex cases such as \languageC-10
and \languageRust-17,
\proj does not exhibit the sharp slowdown observed in
\vulcan and \lpr.
This is because \proj is not limited to repeatedly exploring
the search paths of traditional reducers; instead, its
case-specific, semantic-aware reductions allow the agent to
identify promising edits directly, making the running time
more robust to program size and structure.

\mytightparagraph{Monetary Cost.}
We measure the cost of \llm-assisted reducers,
\ie, \proj and \lpr, by the monetary expense of the \llm API calls
in US dollars (\$).
On average,
\proj spends \$0.035 on the reducer agent and a further
\$0.020 on the reflector agent per case,
against \$0.014 for \lpr.
Although higher than \lpr, we suggest
that the cost is reasonable given \proj achieves
superior reduction effectiveness to \lpr. Moreover,
the reflector's cost is a one-time expense
that will continue to benefit future cases
and saving cost
(see \cref{subsec:generalizability-of-learned-reducer}).

\subsection{RQ2: Generalizability of the Learned Reducer}
\label{subsec:generalizability-of-learned-reducer}

\begin{figure}[!h]
    \centering
    \includegraphics[width=\columnwidth]{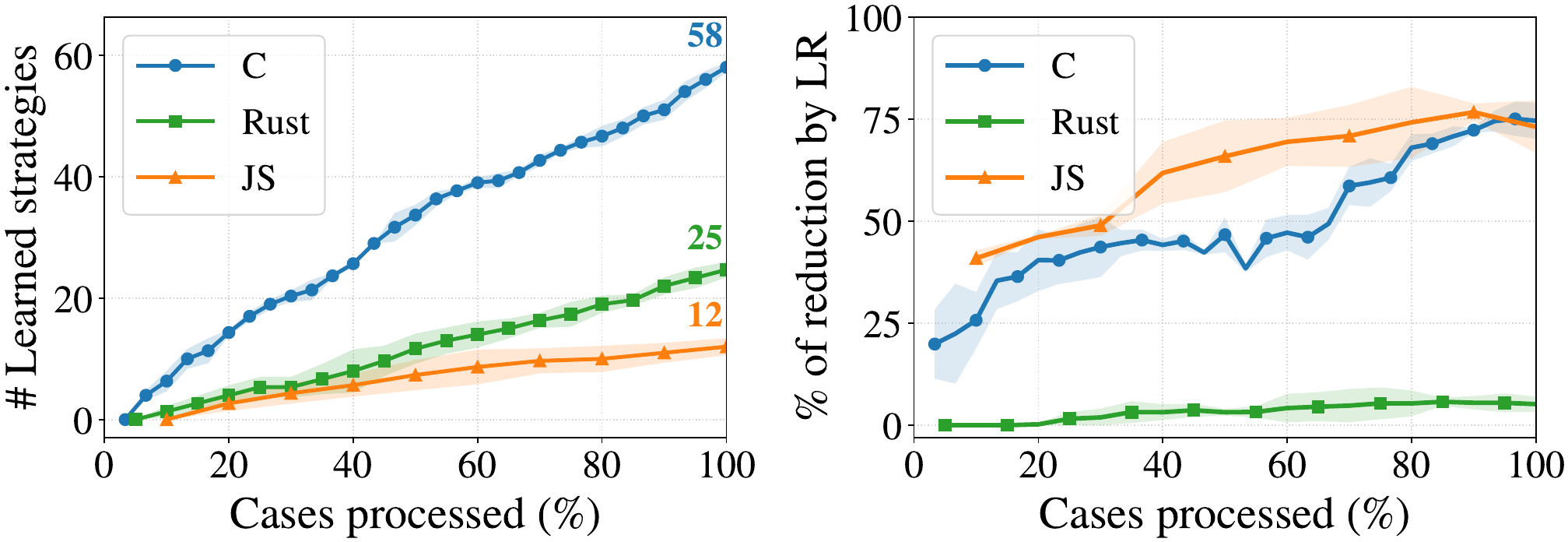}
    \vspace{-0.5cm}
    \caption{
        \small
        The learned reducer (LR) during RQ1: accumulated
        strategies (left) and their share of the reduction
        (right) as cases are processed.
        Lines show the mean over three runs, and shaded bands
        indicate one standard deviation.
    }
    \label{fig:learned_reducer_trend}
\end{figure}

\begin{figure*}[t]
  \centering
  \footnotesize
  \setlength{\tabcolsep}{5pt}
  \renewcommand{\arraystretch}{0.82}
  \begin{tabular}{c@{\hspace{4pt}}|@{\hspace{6pt}}p{0.26\textwidth} p{0.39\textwidth} p{0.252\textwidth}}
  \toprule
  & \textbf{Strategy} & \textbf{Pattern} & \textbf{Rewrite} \\
  \midrule
  \multirow{4}{*}{\rotatebox[origin=c]{90}{C}}
  & \texttt{remove\_unused\_param} & \texttt{f(\dots,p,\dots)} with calls \texttt{f(\dots,a,\dots)}, \emph{$p$ unused} & drop \texttt{p} and the matching \texttt{a} at every call \\
  & \texttt{change\_return\_type\_to\_void} & \texttt{T f(\dots)\{\dots return e;\dots\}}, \emph{no caller uses the
  result} & \texttt{void f(\dots)\{\dots return;\dots\}} \\
  & \texttt{inline\_forwarding\_wrapper} & \texttt{T f(x)\{ return g(x); \}} with calls \texttt{f(a)} & remove
  \texttt{f}; rewrite \texttt{f(a)} to \texttt{g(a)} \\
  & \texttt{inline\_predicate\_function} & \texttt{int f(p)\{ if(C) return 1; \}} with calls \texttt{f(a)} & remove
  \texttt{f}; rewrite \texttt{f(a)} to \texttt{C} with $p\!\mapsto\!a$ \\
  \midrule
  \multirow{2}{*}{\rotatebox[origin=c]{90}{Rust}}
  & \texttt{match\_wildcard\_continue\_to\_if} & \texttt{match e \{L => B, \_ => continue\}}, \emph{inside a loop} &
  \texttt{if e == L \{ B \}} \\
  & \texttt{inline\_single\_use\_let} & \texttt{let x = e; \dots use(x)\dots}, \emph{x used once} & \texttt{\dots
  use(e)\dots} \\
  \midrule
  \multirow{2}{*}{\rotatebox[origin=c]{90}{JS}}
  & \texttt{new\_function\_to\_eval} & \texttt{new Function(s)}, \emph{called with no arguments} & \texttt{eval(s)} \\
  & \texttt{reflect\_get\_to\_bracket} & \texttt{Reflect.get(o, k)} & \texttt{o[k]} \\
  \bottomrule
  \end{tabular}
  \caption{
    Example strategies in the learned reducer distilled by the reflector agent.
  }
  \label{fig:strategy-examples}
  \end{figure*}

We evaluate the generalizability of the learned reducer
from two complementary angles. First, we analyze the learned
reducer's contribution \emph{during} the RQ1 experiments.
Since benchmarks are processed sequentially with the
reflector agent enabled, the learned reducer grows
with accumulated strategies throughout the experiment.
\cref{fig:learned_reducer_trend} tracks, for each language,
both the number of strategies the learned reducer
accumulates and the share of the reduction they contribute
as cases are processed in sequence.
By the end of RQ1, the learned reducer accumulates
58, 25, and 12 strategies for \languageC, \languageRust,
and \languageJS, respectively.
\cref{fig:strategy-examples} provides several examples
of strategies for each language.
The learned reducer accumulates strategies steadily
throughout the run, so each new program is increasingly
reduced by this deterministic prelude before the
reducer agent ever sees it.
For \languageC and \languageJS the contribution is
substantial, reaching around 46\% and 52\% of the total
token reduction: patterns distilled from earlier cases apply
automatically to later ones, progressively offloading
the reducer agent. The effect is weaker for \languageRust.
Although the reflector accumulates a fair number of Rust
strategies, they contribute only 5\% of its reduction,
which we attribute to the heterogeneity of the
\languageRust benchmarks: patterns distilled from one
program recur far less often in the others, so few
strategies apply beyond the case that produced them.

Furthermore, we assess the generalizability of the
learned reducer on \emph{entirely unseen} programs.
We freeze the learned reducer at the state accumulated
after RQ1 completes on \benchmarkCTrain,
and apply it to the 30 held-out programs in \benchmarkCTest.
We then measure the token reduction achieved by
the learned reducer alone on these unseen programs.
This experiment directly tests whether the distilled
strategies apply to new cases without any
reducer agent intervention.
We conduct this held-out evaluation exclusively on
\languageC, as \languageRust and \languageJS benchmarks
contain too few programs for a statistically meaningful
train/test split.

\begin{figure}[!htbp]
    \raggedright
    \hspace*{-0.1cm}%
    \begin{subfigure}{0.495\columnwidth}
        \includegraphics[width=\columnwidth]{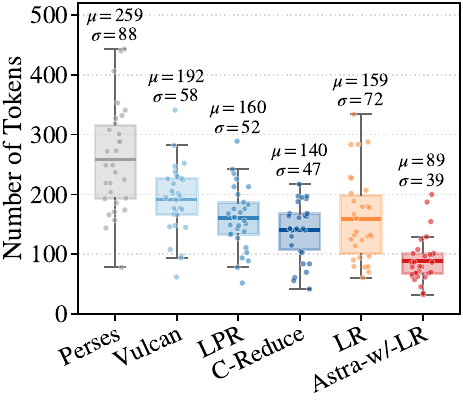}
        \captionsetup{margin={0.5cm, 0cm},skip=0pt}
        \caption{
            \small    
            Effectiveness
        }
        \label{subfig:learn_reducer_effectiveness_box}
    \end{subfigure}
    \begin{subfigure}{0.495\columnwidth}
        \includegraphics[width=\columnwidth]{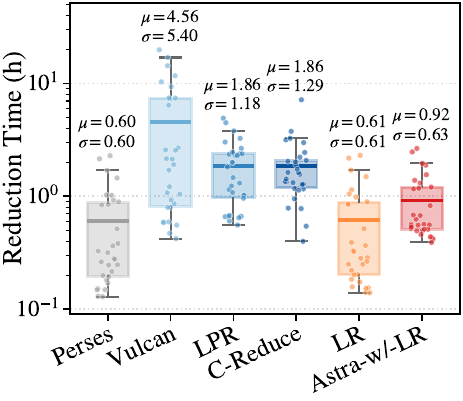}
        \captionsetup{margin={0.5cm, 0cm},skip=0pt}
        \caption{
            \small
            Efficiency
        }
        \label{subfig:learn_reducer_efficiency_box}
    \end{subfigure}
    \vspace{-0.6cm}
    \caption{
        Evaluation results of the learned reducer (LR),
        \proj with the learned reducer (\proj-w/-LR),
        and the baselines on
        30 previously unseen 
        \languageC programs.
    }
    \label{fig:learned_reducer_box}
\end{figure}

\cref{fig:learned_reducer_box} reports the effectiveness
(\cref{subfig:learn_reducer_effectiveness_box}) and efficiency
(\cref{subfig:learn_reducer_efficiency_box}) of the frozen
learned reducer on the 30 held-out programs.
The learned reducer alone shrinks these unseen programs to
159 tokens on average. This already surpasses \vulcan
(192 tokens) and \lpr (160 tokens),
which relies on aggressive hand-designed transformations,
and approaches \creduce (140 tokens), a far heavier machinery.
The distilled strategies thus
generalize well beyond the programs that produced them:
\emph{\proj captures genuinely reusable reduction knowledge,
not case-specific edits.}
Moreover, this effectiveness comes at marginal additional cost.
Running with no \llm, the learned reducer adds only tens of
seconds on top of the \perses prelude it builds on; the full
deterministic pipeline reaches these sizes in roughly 0.61
hours per program, well below the 4.6 and 1.9 hours that \vulcan
and \creduce require on average. In effect, \proj distills,
automatically and without any expert effort, a reducer that
rivals years of hand-engineered tooling at a fraction of the
time. Moreover, the full \proj pipeline further reduces
the programs to 89 tokens on average, demonstrating
the value of the agent's case-specific reasoning
that complements the learned reducer.

\begin{figure}[!htbp]
    \centering
    \hspace{-14pt}
    \includegraphics[width=0.8\columnwidth]{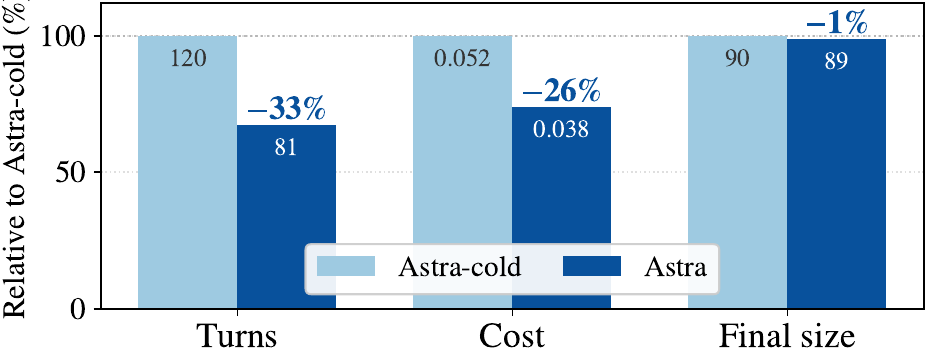}
    \vspace{-0.2cm}
    \caption{
        \small
        Comparison of \proj with and without the learned reducer.
    }
    \label{fig:astra_vs_no_learned_reducer}
\end{figure}

Furthermore, we compare \proj (with the learned reducer)
with \projCold, which represents a cold start version of \proj
without the learned reducer.
As shown in \cref{fig:astra_vs_no_learned_reducer},
when reducing the same 30 programs, \proj requires
\astraVsAstraColdLLMTurns less \llm API calls and
\astraVsAstraColdCost less cost than \projCold,
while achieving comparable reduction effectiveness.
This indicates that
the learned reducer absorbs much of the reduction with cheap,
deterministic strategies, offloading workload from
the reducer agent, thus saving the cost of \llm API calls.

We believe these numbers understate the true potential of the
learned reducer. Its strategies are distilled from only 30
training programs, and \cref{fig:learned_reducer_trend} shows
the strategy set still growing with no sign of saturation. As
\proj processes more cases, we therefore expect the learned
reducer to keep improving on its own, recovering an increasing
share of the reduction with no \llm at inference time.

\subsection{RQ3: Ablation Study}
\label{subsec:ablation-study}

This section investigates the contribution of each component and
design choice in \proj.

\mytightparagraph{Two-Level Loop Structure.}
The reducer agent is the core component of \proj.
We first verify the necessity of the two-level loop structure
in the reducer agent.
To achieve this,
we compare the default configuration
($I_{\max} = 3$ iterations with
$A_{\max} = \maxTurnsFirstIteration,
\maxTurnsSecondIteration,
\maxTurnsThirdIteration$
attempts) against two variants that remove one level of
the loop structure.
The first variant removes the outer-loop reset by using
$I_{\max} = 1$ and $A_{\max} = \maxTurnsTotal$,
keeping all attempts within a single \llm session.
The second variant removes the inner-loop continuity by using
$I_{\max} = \maxTurnsTotal$ and $A_{\max} = 1$,
starting a fresh \llm session after every attempt.
All three configurations consume the same total budget of
\maxTurnsTotal attempts, ensuring a fair comparison.

\begin{figure}[!h]
    \centering
    \hspace{-15pt}
    \includegraphics[width=0.8\columnwidth]{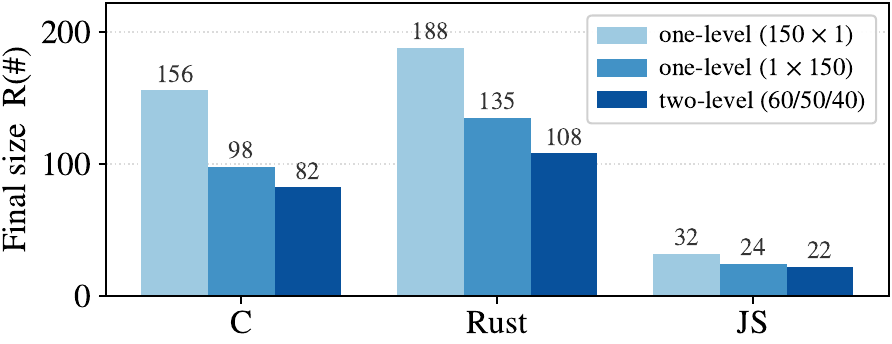}
    \vspace{-0.25cm}
    \caption{
        Ablation study on the two-level loop structure design.
    }
    \label{fig:two_level_loop_ablation}
\end{figure}

The results are shown in \cref{fig:two_level_loop_ablation},
where the default two-level loop structure achieves the most
effective reduction on all three benchmark suites,
demonstrating that both levels are necessary.
Removing the inner-loop continuity hurts the most:
with a fresh session after every attempt,
the agent loses track of what it has already tried and
wastes much of its budget re-proposing edits the
property checker has already rejected.
Within-session context lets each attempt
learn from past failures.
The outer loop helps too, as a single uninterrupted session
eventually stalls as the agent fixates on its earlier
reasoning, whereas periodically resetting it to the
current best program lets the agent push the reduction further.

\mytightparagraph{Two Reduction Modes.}
We analyze the respective contributions of the two
reduction modes, \ie, default mode and exploration mode,
to the overall reduction.
We measure the usage frequency and
the contribution of each mode during the reduction.
As shown in \cref{tab:reduction-mode-ablation},
the default mode does the bulk of the work,
contributing the large majority of the token
reduction across all three suites, while exploration
is used typically more sparingly.
This is consistent with our design, where the default
mode is the primary driver and exploration plays a
tactical, supporting role.
Since an exploration move never lowers the best
program on its own, we credit it only with the
reductions it directly enables, so its contribution
stays below its usage frequency, as many exploratory
detours are reverted once the budget is spent.
It is nonetheless meaningful: the reductions it
unlocks, such as deleting a function that inlining
has just made dead, are ones a strictly
size-decreasing reducer would otherwise miss.

\begin{table}[t]
    \centering
    \caption{
        \small
        Frequency and contribution of two reduction modes.
    }
    \label{tab:reduction-mode-ablation}
    \setlength{\tabcolsep}{8pt}
    \renewcommand{\arraystretch}{0.9}
    \begin{tabular}{c cc cc}
      \toprule
      \multirow{2}{*}{Suite} & \multicolumn{2}{c}{\textbf{Frequency}} & \multicolumn{2}{c}{\textbf{Contribution}} \\
      \cmidrule(lr){2-3}\cmidrule(lr){4-5}
                             & Default & Explore & Default & Explore \\
      \midrule
      C    & 70\% & 30\% & 86\% & 14\% \\
      Rust & 72\% & 28\% & 87\% & 13\% \\
      JS   & 49\% & 51\% & 80\% & 20\% \\
      \bottomrule
    \end{tabular}
  \end{table}

\mytightparagraph{The Effect of \llm Models.}
We study the effect of different \llm models
on the performance of \proj.
Specifically,
besides the default model \deepseekVfourflash,
we additionally evaluate \proj on
\mimoVTwoPointFive~\cite{mimo2026v25} and
\miniMaxMThree~\cite{minimax2026m3}.
As shown in \cref{tab:model-and-harness-ablation},
\proj is largely robust to the choice of \llm backend:
across the three models, the average reduced size stays
within a narrow band on every language.
This suggests that the effectiveness of \proj
rests primarily on the reduction method rather than on
any single model. While a stronger reasoning model may push
the reduction further, we leave a closer study
of this effect to future work.
\begin{table}[!h]
    \centering
    \caption{
        Ablation study on \llm models and agent harness.
    }
    \label{tab:model-and-harness-ablation}
    \setlength{\tabcolsep}{10pt}
    \renewcommand{\arraystretch}{0.9}
    \begin{tabular}{cccc}
      \toprule
      Configuration & C & Rust & JS\\
      \midrule
      \proj-{\scriptsize \deepseekVfourflash} & 80.1 & \textbf{112.2} & \textbf{21.5} \\
      \proj-{\scriptsize \mimoVTwoPointFive} & 77.1 & 116.2 & 24.5 \\
      \proj-{\scriptsize \miniMaxMThree} & \textbf{70.4} & 112.5 & 23.0 \\
      \miniSweAgent & 152.7 & 177.5 & 25.2 \\
      \aider & 126.9 & 169.9 & 26.8 \\
      \bottomrule
    \end{tabular}
  \end{table}

\mytightparagraph{The Effect of Agent Harness.}
An \llm agent is conceptually a combination of the \llm
model and the harness. From this perspective,
we further assess the effect of the harness layer
by comparing \proj, our specialized program reduction agent,
with \miniSweAgent~\cite{swe-agent} and \aider~\cite{aider},
two popular general-purpose SWE agents.
We disable the learned reducer in \proj for a fair comparison.
We equip \miniSweAgent and \aider with the same model
as \proj, \ie, \deepseekVfourflash,
and instruct them with a well-defined prompt
to perform program reduction.
The input programs are pruned by \perses
before being fed to the agents.
As shown in \cref{tab:model-and-harness-ablation},
with the harness as the only variable, \proj produces
substantially smaller programs than both general agents.
This isolates the contribution
of the harness: a general agent left to free-form edits tends
to stall on large, structured programs, whereas \proj's
specialized two-level reduction loop systematically explores
and removes redundant code that the general agents leave in place.
The harness, not the model alone, is what turns an \llm into
an effective program reducer.

\section{Discussion}
\label{sec:discussion}

\mytightparagraph{Threats to Validity.}
The primary threat is \llm non-determinism, which we mitigate
by repeating each experiment three times and reporting the mean.
Another threat is potential data leakage: \llms may have
seen some benchmarks during pre-training and reproduce
memorized reductions rather than genuinely reasoning about them.
To mitigate this, first, our prompts include no
bug identifiers or metadata linked to a known bug.
Second, our suites include 40 programs generated by fuzzers
that have never been publicly released, so the \llm could not have
seen them during training.

\mytightparagraph{Limitations and Future Work.}
A potential issue is \emph{strategy explosion}:
although the strategies in the learned reducer remain controllable
in our experiments, and introduce negligible overhead,
the pool grows with every case and could, over a long deployment,
accumulate a large number of narrow, rarely-used strategies,
increasing runtime overhead and maintenance costs.
We may then need mechanisms to govern the growing pool.
Promising directions include retiring low-yield strategies based on
how often they fire and how much they save, and prioritizing the most
productive ones so that runtime overhead stays low as the
pool grows. We leave this to future work.

\section{Related Work}
\label{sec:related-work}

\proj is most directly related to prior work on program reduction.
Delta Debugging~\cite{delta-debugging} is the cornerstone of program
reduction. Its core algorithm, \ddmin, minimizes list-structured
inputs and underpins a large family of subsequent
reducers~\cite{perses,vulcan,hdd,xu2025boosting,ddsmt,hddr,ppr}.
Follow-up work replaces \ddmin's fixed strategy with
probabilistic~\cite{probdd}, counter-based~\cite{cdd},
or weighted~\cite{wdd} models for faster convergence.
Structure-aware reducers apply \ddmin across parse-tree
levels~\cite{hdd} or achieve fully syntax-guided
reduction~\cite{perses,adhoc}. \vulcan further pushes
minimality of reduction with aggressive transformations~\cite{vulcan}.
Other reducers operate at the token level~\cite{trec} or
speed up search via caching~\cite{cache}.
Language-specific reducers tailor strategies to a single
language, \eg, C~\cite{creduce}, SMT-LIBv2~\cite{ddsmt},
or Java bytecode~\cite{j-reduce}.
Closest to our work, \lpr~\cite{lpr} pioneered LLM-assisted
reduction but restricts the \llm to predefined transformations.
\proj instead lets agentic \llms reason freely about each
program and distills successful reductions into reusable
strategies.

\llms are increasingly applied to compiler engineering.
The most active front is \emph{compiler
testing}~\cite{metamut,whitefox,legofuzz,gu2023llm,kitten,huangoptfuzz},
where \llms generate programs that exercise compilers in ways
hand-crafted fuzzers rarely reach, \eg,
WhiteFox~\cite{whitefox} synthesizes inputs targeting specific
optimizations.
\llms have also been used to improve compiler
optimizations~\cite{lpo,peepgen,taneja2025llm}, \eg,
LPO~\cite{lpo} discovers missed peephole optimizations with \llms.
Recently, Zheng \etal~\cite{zheng2026agentic} studied
\llm agents for fixing compiler bugs.
In the same agentic spirit, \proj brings agents to program
reduction.

\section{Conclusion}
\label{sec:conclusion}
This paper reimagines program reduction as a semantic-aware
exploratory reasoning process, and realizes this vision in
\proj, an agentic framework that pairs a \emph{reducer agent}
with a \emph{reflector agent} for program reduction.
\proj reduces programs beyond the reach of any fixed
transformation set, and, unlike prior reducers, it does not stand
still, but continuously improves as it accumulates experience.
\proj thus establishes a new paradigm for program reduction,
one in which a reducer no longer merely applies fixed rules but
reasons about each program and remembers what it learns.

\newpage

\bibliographystyle{IEEEtran}
\bibliography{citations}

\begin{thebibliography}{10}
\providecommand{\url}[1]{#1}
\csname url@samestyle\endcsname
\providecommand{\newblock}{\relax}
\providecommand{\bibinfo}[2]{#2}
\providecommand{\BIBentrySTDinterwordspacing}{\spaceskip=0pt\relax}
\providecommand{\BIBentryALTinterwordstretchfactor}{4}
\providecommand{\BIBentryALTinterwordspacing}{\spaceskip=\fontdimen2\font plus
\BIBentryALTinterwordstretchfactor\fontdimen3\font minus \fontdimen4\font\relax}
\providecommand{\BIBforeignlanguage}[2]{{%
\expandafter\ifx\csname l@#1\endcsname\relax
\typeout{** WARNING: IEEEtran.bst: No hyphenation pattern has been}%
\typeout{** loaded for the language `#1'. Using the pattern for}%
\typeout{** the default language instead.}%
\else
\language=\csname l@#1\endcsname
\fi
#2}}
\providecommand{\BIBdecl}{\relax}
\BIBdecl

\bibitem{emi}
V.~Le, M.~Afshari, and Z.~Su, ``Compiler validation via equivalence modulo inputs,'' \emph{ACM Sigplan Notices}, vol.~49, no.~6, pp. 216--226, 2014.

\bibitem{csmith}
X.~Yang, Y.~Chen, E.~Eide, and J.~Regehr, ``Finding and understanding bugs in c compilers,'' in \emph{Proceedings of the 32nd ACM SIGPLAN conference on Programming language design and implementation}, 2011, pp. 283--294.

\bibitem{live-code-mutation}
C.~Sun, V.~Le, and Z.~Su, ``Finding compiler bugs via live code mutation,'' in \emph{Proceedings of the 2016 ACM SIGPLAN international conference on object-oriented programming, systems, languages, and applications}, 2016, pp. 849--863.

\bibitem{yarpgen}
V.~Livinskii, D.~Babokin, and J.~Regehr, ``Random testing for c and c++ compilers with yarpgen,'' \emph{Proceedings of the ACM on Programming Languages}, vol.~4, no. OOPSLA, pp. 1--25, 2020.

\bibitem{lidbury2015many}
C.~Lidbury, A.~Lascu, N.~Chong, and A.~F. Donaldson, ``Many-core compiler fuzzing,'' \emph{ACM SIGPLAN Notices}, vol.~50, no.~6, pp. 65--76, 2015.

\bibitem{gccbugreport}
\BIBentryALTinterwordspacing
GCC-Wiki. (2026) A guide to testcase reduction. [Online]. Available: \url{https://gcc.gnu.org/wiki/A_guide_to_testcase_reduction}
\BIBentrySTDinterwordspacing

\bibitem{llvmbugreport}
\BIBentryALTinterwordspacing
LLVM. (2026) How to submit an llvm bug report. [Online]. Available: \url{https://llvm.org/docs/HowToSubmitABug.html}
\BIBentrySTDinterwordspacing

\bibitem{jerrybugreport}
\BIBentryALTinterwordspacing
JerryScript. (2026) Bug report. [Online]. Available: \url{https://github.com/jerryscript-project/jerryscript/blob/master/.github/ISSUE_TEMPLATE/bug_report.md}
\BIBentrySTDinterwordspacing

\bibitem{cpythonreport}
\BIBentryALTinterwordspacing
CPython. (2026) Bug report. [Online]. Available: \url{https://github.com/python/cpython/issues/new?assignees=&labels=type-bug&template=bug.md}
\BIBentrySTDinterwordspacing

\bibitem{creduce}
J.~Regehr, Y.~Chen, P.~Cuoq, E.~Eide, C.~Ellison, and X.~Yang, ``Test-case reduction for c compiler bugs,'' in \emph{Proceedings of the 33rd ACM SIGPLAN conference on Programming Language Design and Implementation}, 2012, pp. 335--346.

\bibitem{perses}
C.~Sun, Y.~Li, Q.~Zhang, T.~Gu, and Z.~Su, ``Perses: Syntax-guided program reduction,'' in \emph{Proceedings of the 40th International Conference on Software Engineering}, 2018, pp. 361--371.

\bibitem{vulcan}
Z.~Xu, Y.~Tian, M.~Zhang, G.~Zhao, Y.~Jiang, and C.~Sun, ``Pushing the limit of 1-minimality of language-agnostic program reduction,'' \emph{Proceedings of the ACM on Programming Languages}, vol.~7, no. OOPSLA1, pp. 636--664, 2023.

\bibitem{xu2025boosting}
Z.~Xu, Y.~Tian, M.~Zhang, and C.~Sun, ``Boosting program reduction with the missing piece of syntax-guided transformations,'' \emph{Proceedings of the ACM on Programming Languages}, vol.~9, no. OOPSLA2, pp. 86--112, 2025.

\bibitem{trec}
Z.~Xu, Y.~Tian, M.~Zhang, J.~Zhang, P.~Liu, Y.~Jiang, and C.~Sun, ``T-rec: Fine-grained language-agnostic program reduction guided by lexical syntax,'' \emph{ACM Transactions on Software Engineering and Methodology}, vol.~34, no.~2, pp. 1--31, 2025.

\bibitem{lpr}
M.~Zhang, Y.~Tian, Z.~Xu, Y.~Dong, S.~H. Tan, and C.~Sun, ``Lpr: Large language models-aided program reduction,'' in \emph{Proceedings of the 33rd ACM SIGSOFT International Symposium on Software Testing and Analysis}, 2024, pp. 261--273.

\bibitem{llvm-198755}
\BIBentryALTinterwordspacing
{LLVM Project}. (2026) {[LLVM][RISCV][RVV] MachineVerifier failure ``Virtual register defs don't dominate all uses'' with RVV vector stores at -O2/-O3 -flto}. GitHub issue \#198755. [Online]. Available: \url{https://github.com/llvm/llvm-project/issues/198755}
\BIBentrySTDinterwordspacing

\bibitem{gcc-125538}
\BIBentryALTinterwordspacing
{GCC Bugzilla community}. (2026) {GCC Bugzilla Report 125538: gfortran ICE with -std=f2008 and associates}. Bugzilla report \#125538. [Online]. Available: \url{https://gcc.gnu.org/bugzilla/show_bug.cgi?id=125538}
\BIBentrySTDinterwordspacing

\bibitem{latra}
Z.~Xu, Y.~Wang, Y.~Tian, M.~Zhang, and C.~Sun, ``Latra: A template-based language-agnostic transformation framework for effective program reduction,'' in \emph{2025 40th IEEE/ACM International Conference on Automated Software Engineering (ASE)}.\hskip 1em plus 0.5em minus 0.4em\relax IEEE, 2025, pp. 2274--2285.

\bibitem{yao2022react}
S.~Yao, J.~Zhao, D.~Yu, N.~Du, I.~Shafran, K.~Narasimhan, and Y.~Cao, ``React: Synergizing reasoning and acting in language models,'' \emph{arXiv preprint arXiv:2210.03629}, 2022.

\bibitem{llvm27747}
\BIBentryALTinterwordspacing
L.~Bugzilla. (2016) Bug 27747 - clang crashes on valid code at -o1 and above on x86\_64-linux-gnu. [Online]. Available: \url{https://bugs.llvm.org/show_bug.cgi?id=27747}
\BIBentrySTDinterwordspacing

\bibitem{shinn2023reflexion}
N.~Shinn, F.~Cassano, A.~Gopinath, K.~Narasimhan, and S.~Yao, ``Reflexion: Language agents with verbal reinforcement learning,'' \emph{Advances in neural information processing systems}, vol.~36, pp. 8634--8652, 2023.

\bibitem{wang2023voyager}
G.~Wang, Y.~Xie, Y.~Jiang, A.~Mandlekar, C.~Xiao, Y.~Zhu, L.~Fan, and A.~Anandkumar, ``Voyager: An open-ended embodied agent with large language models,'' \emph{arXiv preprint arXiv:2305.16291}, 2023.

\bibitem{deepseek-v4-flash}
Y.~Wang, Q.~Zhang, J.~Yu, T.~Liang, D.~Ma, X.~Hu, Z.~Lin, C.~Li, Z.~Wang, J.~Li \emph{et~al.}, ``Flashmemory-deepseek-v4: Lightning index ultra-long context via lookahead sparse attention,'' \emph{arXiv preprint arXiv:2606.09079}, 2026.

\bibitem{mimo2026v25}
{Xiaomi MiMo Team}, ``Mimo-v2.5: Native omnimodal mixture-of-experts model with agentic capabilities,'' \url{https://huggingface.co/XiaomiMiMo/MiMo-V2.5}, 2026.

\bibitem{minimax2026m3}
{MiniMax AI}, ``Minimax-m3: Large-scale multimodal foundation model,'' \url{https://www.minimax.io/blog/minimax-m3}, 2026.

\bibitem{swe-agent}
J.~Yang, C.~Jimenez, A.~Wettig, K.~Lieret, S.~Yao, K.~Narasimhan, and O.~Press, ``Swe-agent: Agent-computer interfaces enable automated software engineering,'' \emph{Advances in Neural Information Processing Systems}, vol.~37, pp. 50\,528--50\,652, 2024.

\bibitem{aider}
P.~Gauthier, ``Aider: Ai pair programming in your terminal,'' \url{https://github.com/aider-ai/aider}, 2025, gitHub repository.

\bibitem{delta-debugging}
A.~Zeller and R.~Hildebrandt, ``Simplifying and isolating failure-inducing input,'' \emph{IEEE Transactions on software engineering}, vol.~28, no.~2, pp. 183--200, 2002.

\bibitem{hdd}
G.~Misherghi and Z.~Su, ``Hdd: hierarchical delta debugging,'' in \emph{Proceedings of the 28th international conference on Software engineering}, 2006, pp. 142--151.

\bibitem{ddsmt}
A.~Niemetz and A.~Biere, ``ddsmt: a delta debugger for the smt-lib v2 format,'' in \emph{Proceedings of the 11th International Workshop on Satisfiability Modulo Theories, SMT}, 2013, pp. 8--9.

\bibitem{hddr}
{\'A}.~Kiss, R.~Hodov{\'a}n, and T.~Gyim{\'o}thy, ``Hddr: a recursive variant of the hierarchical delta debugging algorithm,'' in \emph{Proceedings of the 9th ACM SIGSOFT International Workshop on Automating TEST Case Design, Selection, and Evaluation}, 2018, pp. 16--22.

\bibitem{ppr}
M.~Zhang, Z.~Xu, Y.~Tian, Y.~Jiang, and C.~Sun, ``Ppr: Pairwise program reduction,'' in \emph{Proceedings of the 31st ACM Joint European Software Engineering Conference and Symposium on the Foundations of Software Engineering}, 2023, pp. 338--349.

\bibitem{probdd}
G.~Wang, R.~Shen, J.~Chen, Y.~Xiong, and L.~Zhang, ``Probabilistic delta debugging,'' in \emph{Proceedings of the 29th ACM joint meeting on european software engineering conference and symposium on the foundations of software engineering}, 2021, pp. 881--892.

\bibitem{cdd}
M.~Zhang, Z.~Xu, Y.~Tian, X.~Cheng, and C.~Sun, ``Toward a better understanding of probabilistic delta debugging,'' in \emph{2025 IEEE/ACM 47th International Conference on Software Engineering (ICSE)}.\hskip 1em plus 0.5em minus 0.4em\relax IEEE, 2025, pp. 2024--2035.

\bibitem{wdd}
X.~Zhou, Z.~Xu, M.~Zhang, Y.~Tian, and C.~Sun, ``Wdd: Weighted delta debugging,'' in \emph{2025 IEEE/ACM 47th International Conference on Software Engineering (ICSE)}, 2025, pp. 1592--1603.

\bibitem{adhoc}
J.~L. Tian, M.~Zhang, Z.~Xu, Y.~Tian, Y.~Dong, and C.~Sun, ``Ad hoc syntax-guided program reduction,'' in \emph{Proceedings of the 31st ACM Joint European Software Engineering Conference and Symposium on the Foundations of Software Engineering}, 2023, pp. 2137--2141.

\bibitem{cache}
Y.~Tian, X.~Zhang, Y.~Dong, Z.~Xu, M.~Zhang, Y.~Jiang, S.-C. Cheung, and C.~Sun, ``On the caching schemes to speed up program reduction,'' \emph{ACM Transactions on Software Engineering and Methodology}, vol.~33, no.~1, pp. 1--30, 2023.

\bibitem{j-reduce}
C.~G. Kalhauge and J.~Palsberg, ``Binary reduction of dependency graphs,'' in \emph{Proceedings of the 2019 27th ACM Joint Meeting on European Software Engineering Conference and Symposium on the Foundations of Software Engineering}, 2019, pp. 556--566.

\bibitem{metamut}
X.~Ou, C.~Li, Y.~Jiang, and C.~Xu, ``The mutators reloaded: Fuzzing compilers with large language model generated mutation operators,'' in \emph{Proceedings of the 29th ACM International Conference on Architectural Support for Programming Languages and Operating Systems, Volume 4}, 2024, pp. 298--312.

\bibitem{whitefox}
C.~Yang, Y.~Deng, R.~Lu, J.~Yao, J.~Liu, R.~Jabbarvand, and L.~Zhang, ``Whitefox: White-box compiler fuzzing empowered by large language models,'' \emph{Proceedings of the ACM on Programming Languages}, vol.~8, no. OOPSLA2, pp. 709--735, 2024.

\bibitem{legofuzz}
Y.~Ni and S.~Li, ``Interleaving large language models for compiler testing,'' \emph{Proceedings of the ACM on Programming Languages}, vol.~9, no. OOPSLA2, pp. 815--841, 2025.

\bibitem{gu2023llm}
Q.~Gu, ``Llm-based code generation method for golang compiler testing,'' in \emph{Proceedings of the 31st ACM Joint European Software Engineering Conference and Symposium on the Foundations of Software Engineering}, 2023, pp. 2201--2203.

\bibitem{kitten}
Y.~Xie, Z.~Xu, Y.~Tian, M.~Zhou, X.~Zhou, and C.~Sun, ``Kitten: A simple yet effective baseline for evaluating llm-based compiler testing techniques,'' in \emph{Proceedings of the 34th ACM SIGSOFT International Symposium on Software Testing and Analysis}, 2025, pp. 21--25.

\bibitem{huangoptfuzz}
Y.~Huang, Y.~Yang, M.~Sun, J.~Wu, Q.~Li, Z.~Lu, and Y.~Zhou, ``Optfuzz: Enhancing compiler testing via llm-powered compilation option generation,'' \emph{ACM Transactions on Software Engineering and Methodology}.

\bibitem{lpo}
Z.~Xu, H.~Xu, Y.~Tian, X.~Zhou, and C.~Sun, ``Lpo: Discovering missed peephole optimizations with large language models,'' in \emph{Proceedings of the 31st ACM International Conference on Architectural Support for Programming Languages and Operating Systems, Volume 2}, 2026, pp. 1136--1150.

\bibitem{peepgen}
C.~Liao, H.~Xu, X.~Zhou, Z.~Xu, and C.~Sun, ``Leveraging large language models for generalizing peephole optimizations,'' \emph{arXiv preprint arXiv:2603.18477}, 2026.

\bibitem{taneja2025llm}
J.~Taneja, A.~Laird, C.~Yan, M.~Musuvathi, and S.~K. Lahiri, ``Llm-vectorizer: Llm-based verified loop vectorizer,'' in \emph{Proceedings of the 23rd ACM/IEEE International Symposium on Code Generation and Optimization}, 2025, pp. 137--149.

\bibitem{zheng2026agentic}
Y.~Zheng, C.~Li, S.~Li, Y.~Zhang, and Z.~Su, ``Agentic harness for real-world compilers,'' \emph{arXiv preprint arXiv:2603.20075}, 2026.

\end{thebibliography}

\end{document}